\documentclass[a4paper]{article}
\usepackage[a4paper, total={5.5in, 7.5in}]{geometry}

\usepackage[style=nature, url=false, doi=true, isbn=false, sorting=none, maxcitenames=2, maxbibnames=30, giveninits=true]{biblatex}

\renewbibmacro{in:}{}

\DeclareFieldFormat{pages}{#1}
\usepackage{amsmath}
\usepackage{amssymb}
\usepackage{dirtytalk}
\usepackage{lineno}
\usepackage{authblk}
\usepackage{graphicx, wrapfig, rotating}
\usepackage[font=footnotesize]{caption}
\usepackage{float}
\usepackage{booktabs}
\usepackage{enumitem}
\usepackage{abstract}
\usepackage[colorlinks=true, citecolor=green, linkcolor=black, urlcolor=blue]{hyperref}
\usepackage{tikz}
\usepackage{bm}
\usepackage[edges]{forest}
\usepackage{xr}
\usepackage{csquotes} 
\usepackage{mwe}
\providecommand{\keywords}[1]{\textbf{Keywords:} #1}

\begin{document}

\pagenumbering{arabic}
\date{}
\title{Is VARS more intuitive and efficient than Sobol' indices?}

\author[1,2]{Arnald Puy\thanks{Corresponding author}}
\author[3]{Samuele Lo Piano}
\author[3]{Andrea Saltelli}

\affil[1]{\footnotesize{\textit{Department of Ecology and Evolutionary Biology, M31 Guyot Hall, Princeton University, New Jersey 08544, USA. E-Mail: \href{mailto:apuy@princeton.edu}{apuy@princeton.edu}}}}

\affil[2]{\footnotesize{\textit{Centre for the Study of the Sciences and the Humanities (SVT), University of Bergen, Parkveien 9, PB 7805, 5020 Bergen, Norway.}}}

\affil[3]{\footnotesize{\textit{Open Evidence, Universitat Oberta de Catalunya, Edifici 22@, Universitat Oberta de Catalunya, 08018 Barcelona, Spain.}}}

\maketitle

\begin{abstract}
The Variogram Analysis of Response Surfaces (VARS) has been proposed by Razavi and Gupta as a new comprehensive framework in sensitivity analysis. According to these authors, VARS provides a more intuitive notion of sensitivity and it is much more computationally efficient than Sobol' indices. Here we review these arguments and critically compare the performance of VARS-TO, for total-order index, against the total-order Jansen estimator. We argue that, unlike classic variance-based methods, VARS lacks a clear definition of what an \say{important} factor is, and we show that the alleged computational superiority of VARS does not withstand scrutiny. We conclude that while VARS enriches the spectrum of existing methods for sensitivity analysis, especially for a diagnostic use of mathematical models, it complements rather than replaces classic estimators used in variance-based sensitivity analysis.
\end{abstract}


\section{Introduction}
Sensitivity analysis (SA) explores how uncertainty in the output of a model (numerical or otherwise) can be apportioned to different sources of uncertainty in the model input space \cite{Saltelli2002d} \footnote{This article is part of a SI on \say{Sensitivity analysis for environmental modeling.}}. SA is especially needed when complex models, which often formalize partially known processes and include non-linear relations, are used to guide policies in the real world. This is generally the case of models in the Environmental Sciences domain, e.g. on crop water requirements, water availability under climate change, weather forecasting, surface runoff or precipitation and evaporation processes \cite{Doll2002, Pappenberger2011, Vieux2016, Wang2020}. The uncertainties in these models might be either parametric (i.e. exact values for parameters might be unknown, there might be errors in the measurement) or structural (i.e. lack of knowledge on the underlying processes, multiple ways of modeling the same phenomenon), and their combined effect on the model output should be understood to guarantee a robust inference for policy-making. In this context, SA jointly with uncertainty analysis is regarded as an unavoidable step to ensure the quality of the modeling process \cite{Borgonovo2016, Eker2018, Jakeman2006, Saltelli2019b, Saltelli2020, Tarantola2002a}. 

In SA, as in all fields of computational research, different strategies and methods compete to establish themselves as \say{good}, \say{recommended} or \say{best} practices. While variance-based methods and Sobol' indices are deemed to belong to the class of recommended methods \cite{Saltelli2004b, Saltelli2008}, other approaches have been proposed to complement or overcome their limitations (i.e. entropy-based methods \cite{Liu2006a}, the $\delta$ measure \cite{Borgonovo2007}, the Kuiper' metric  \cite{Baucells2013}, or the PAWN index \cite{Pianosi2015, Pianosi2018}). One of the most recent competitors is the Variogram Analysis of Response Surfaces (VARS), proposed by \textcite{Razavi2016a, Razavi2016b}.  According to \textit{Google Scholar} and as of November 2020, the two foundational VARS papers have been cited 86 times, and seem to have been especially embraced by Hydrologists and Water Scientists \cite{Jayathilake2020, Lilhare2020, Krogh2017, Jayathilake2020a}.

\textcite{Razavi2016a, Razavi2016b} report that VARS outperforms Sobol' indices in two main aspects:

\begin{enumerate}
\item It provides a more intuitive assessment of sensitivities and the importance of model inputs in determining the model output.

\item It computes the total-order effect with a much higher computational efficiency (up to two orders of magnitude more efficient).
\end{enumerate}

In the present work we explore these results and benchmark VARS against one of the best Sobol' indices estimator, that of \textcite{Jansen1999}'s. Before engaging in the discussion, we briefly recall hereafter some useful formulae needed to understand the two approaches.    

\subsection{Sobol' indices}
The apparatus of variance-based sensitivity indices, described by \textcite{Sobol1993} and extended by \textcite{Homma1996}, is currently considered as the recommended practice in SA \cite{Saltelli2008}. For a model of $k$ factors $f(\bm{x})=(x_1, x_2, ..., x_k)\in \mathbb{R}^k$, the first-order sensitivity index $S_i$ can be written as 
\begin{equation}
S_i=\frac{V_{x_i} \left [ E_{\bm{x}_{\sim i}} (y | x_i) \right ]}{V(y)}\,.
\label{eq:si}
\end{equation}
The inner mean in Equation~\ref{eq:si} is taken over all-factors-but $x_i$ ($\bm{x}_{\sim i}$), while the outer variance is taken over $x_i$. $V(y)$ is the unconditional variance of the output variable $y$. When the factors are independent, $S_i$ can be defined as a first order term in the variance decomposition of $y$:

\begin{equation}
1=\sum_{i=1}^{k}S_i+\sum_{i}\sum_{i<j}S_{ij}+...+S_{1,2,...,k}\,,
\label{eq:decomposition}
\end{equation}

$S_i$ lends itself to be expressed in plain English as \emph{the fractional reduction in the variance of $y$ which would be obtained on average if $x_i$ could be fixed}. This is because 

\begin{equation}
V(y) = V_{x_i} \left [ E_{\bm{x}_{\sim i}} (y | x_i) \right ] + E_{x_i} \left [ V_{\bm{x}_{\sim i}} (y | x_i) \right ]\,.
\label{eq:variance}
\end{equation}

$E_{x_i} \left [ V_{\bm{x}_{\sim i}} (y | x_i) \right ]$ is the average variance that would be left after fixing $x_i$ to a given value in its uncertainty range. For this reason, $V_{x_i} \left [ E_{\bm{x}_{\sim i}} (y | x_i) \right ]$ must be the average reduction in variance as discussed above. While $V_{\bm{x}_{\sim i}} (y | x_i)$ can be greater than $V(y)$, $E_{x_i} \left [ V_{\bm{x}_{\sim i}} (y | x_i) \right ]$ is always smaller than $V(y)$ as per Equation~\ref{eq:variance}.

Another useful variance-based measure is the total-order index $T_i$ \cite{Homma1996}, which measures the first-order effect of a model input jointly with its interactions up to the $k$-th order:

\begin{equation}
T_i=\frac{E_{\bm{x}_{\sim i}}\left [ V_{x_i} ( y |  \bm{x}_{\sim i}) \right ]}{V(y)}\,.
\end{equation}

The index is called \say{total} because it includes all factors in the variance decomposition (see Equation~\ref{eq:decomposition}) that include the index $i$. For instance, for a model with three factors, $T_1=S_1+S_{1,2}+S_{1,3}+S_{1,2,3}$, and likewise for $T_2$ or $T_3$. The meaning of $T_i$ is \textit{the fraction of variance that would remain on average if $x_i$ is left to vary over its uncertainty range while all other factors are fixed}. Note that the theory of variance-based measures is as flexible as to accommodate \say{group} or \say{set} sensitivities. These are simply the first-order effect of a set of factors: if $u$ is the set of factors $(x_1,x_2)$, then $S_u = S_1+S_2+S_{1,2}$.

\subsection{VARS}
VARS is based on variogram analysis to characterize the spatial structure and variability of a given model output across the input space \cite{Razavi2016a, Razavi2016b}. Let us again consider a function of factors $f(\bm{x})=(x_1, x_2, ..., x_k)\in \mathbb{R}^k$. If $\bm{x}_A$ and $\bm{x}_B$ are two generic points separated by a distance $\bm{h}$, then the variogram $\gamma(.)$ is calculated as
\begin{equation}
\gamma(\bm{x}_A-\bm{x}_B) = \frac{1}{2}V \left [y(\bm{x}_A) - y(\bm{x}_B) \right ]\,,
\end{equation}
and the covariogram $C(.)$ as
\begin{equation}
C(\bm{x}_A-\bm{x}_B) = COV \left [y(\bm{x}_A),  y(\bm{x}_B) \right ]\,.
\end{equation}

Note that

\begin{equation}
V \left [y(\bm{x}_A) - y(\bm{x}_B) \right ] = V \left [y(\bm{x}_A) \right ] + V \left [y(\bm{x}_B) \right ] - 2COV \left [ y(\bm{x}_A), y(\bm{x}_B) \right ]\,.
\end{equation}

Given that $V \left [ y(\bm{x}_A) \right ] = V \left [ y(\bm{x}_B) \right ]$, then 

\begin{equation}
\gamma (\bm{x}_A - \bm{x}_B) = V \left [ y(\bm{x}) \right ] - C(\bm{x}_A, \bm{x}_B)\,.
\label{eq:variogram}
\end{equation}

As mentioned, the points $\bm{x}_A$, $\bm{x}_B$ are spaced by a fixed distance, and $V$, $COV$ are the variance and covariance, respectively. Note that $\gamma(.)$ is defined by the interval separating $\bm{x}_A$, $\bm{x}_B$. To make this clearer one can write  $\bm{h}=\bm{x}_A-\bm{x}_B$, with $\bm{h}=h_1,h_2,…,h_n$, so that 
\begin{equation}
\gamma(\bm{h})= \frac{1}{2}E \left [y(\bm{x} + \bm{h}) - y(\bm{x}) \right ]^2\,,
\end{equation}
where the term $E^2$ in the expression of the variance as the expectation of the square minus the square of the expectation, $V(.)=E(.)^2-E^2(.)$, is assumed to be zero. The practical formula for computing a multidimensional variogram is
\begin{equation}
\gamma(\bm{h})=\frac{1}{2N(\bm{h})} \sum \left [ y(\bm{x}_A) - y(\bm{x}_B) \right ] ^ 2\,,
\end{equation}
where the sum is extended to all $N(\bm{h})$ couples of points $\bm{x}_A,\bm{x}_B$ such that their modulo distance $|\bm{x}_A-\bm{x}_B|$ is $\bm{h}$. \textcite{Razavi2016a, Razavi2016b} suggest some integral measures based on variogram $\gamma$, i.e. the integrated variogram $\Gamma(H_i)$:

\begin{equation}
\Gamma(H_i)=\int_{0}^{H_i}\gamma(h_i)dh_i\,,
\end{equation}
and recommend the use of IVARS$_{10}$, IVARS$_{30}$, and IVARS$_{50}$ (computed for $H$ equal to 10\%, 30\%, and 50\% of the factor range respectively) to explore larger fractions of the variation space of the function, with IVARS$_{50}$ corresponding to the entire interval (in variogram analysis, the maximum meaningful range is one half of the factor range \cite{Cressie2015}). 

Of important practical use, as we shall see, is the directional variogram along one of the axes of the factors space,    

\begin{equation}
\gamma(h_i) = \frac{1}{2}E(y(x_1,...,x_{i+1} + h_i,...,x_n) - y(x_1,...,x_i,...,x_n))^2\,,
\end{equation}
which is evidently computed on all couples of points spaced $h_i$  along the $x_i$ axis, with all other factors being kept fixed. Note that the difference in parentheses is what is called in \textcite{Saltelli2010a} a step along the $x_i$ direction, which is fungible to compute the total sensitivity index $T_i$.  

The equivalent of Equation~\ref{eq:variogram} for the case of the unidirectional variogram $\gamma(h_i)$ is 

\begin{equation}
\gamma_{x^*_{\sim_i}}(h_i)=V(y|x^*_{\sim_i})-C_{x^*_{\sim_i}}(h_i)\,,
\label{eq:uni_variogram}
\end{equation}
where $x^*_{\sim i}$ is a fixed point in the space of non-$x_i$.

In order for VARS to compute the total-order index $T_i$ (labeled as VARS-TO by \textcite{Razavi2016a}), the authors suggest taking the mean value across the factors' space on both sides of Equation~\ref{eq:uni_variogram}, thus obtaining

\begin{equation}
E_{x^*_\sim{i}} \left [ \gamma_{x^*_\sim i}(h_i) \right ]=E_{x^*_\sim{i}}  \left [ V(y|x^*_{\sim i}) \right ] -E_{x^*_\sim{i}}  \left [ C_{x^*_\sim i}(h_i) \right ]\,,
\end{equation}
which can also be written as

\begin{equation}
E_{x^*_\sim{i}} \left [ \gamma_{x^*_\sim i}(h_i) \right ]=V(y)T_i -E_{x^*_\sim{i}}  \left [ C_{x^*_\sim i}(h_i) \right ]\,,
\end{equation}
and therefore

\begin{equation}
T_i=\mbox{VARS-TO}=\frac{E_{x^*_\sim i}\left [ \gamma_{x^*_\sim i}(h_i)\right] + E_{x^*_\sim i} \left [ C_{x^*_\sim i}(h_i) \right ] }{V(y)}\,.
\label{eq:SM_VARS_ti}
\end{equation}

\vspace{10mm}

\section{The issue of intuitivity and importance}
In a paper immediately preceding VARS, \textcite{Razavi2015} already stressed two main drawbacks of global sensitivity analysis: 

\begin{enumerate}
\item The incapacity of variance-based Sobol' indices to appraise the spatial distribution of the model response.
\item The dependence of the \textcite{Morris1991} approach on the step size defined by the analyst, which can significantly condition the final sensitivity value.
\end{enumerate}

VARS was presented as a comprehensive approach which overcomes these drawbacks by encapsulating, in a single sensitivity framework, a ``unified assessment of local and global sensitivity'' \cite[p.~3090]{Razavi2015}. The fact that integrated variogram measures such as IVARS$_{10}$, IVARS$_{30}$ and IVARS$_{50}$ are able to differentiate sensitivities as a function of scale $H$, whereas Sobol' indices do not, is taken as proof of the limitations of the latter. According to \textcite[pp.~427--428, 433--434]{Razavi2016a}, this endows VARS with a more intuitive appraisal of sensitivities.

\begin{figure}[ht]
\centering
\includegraphics[keepaspectratio]{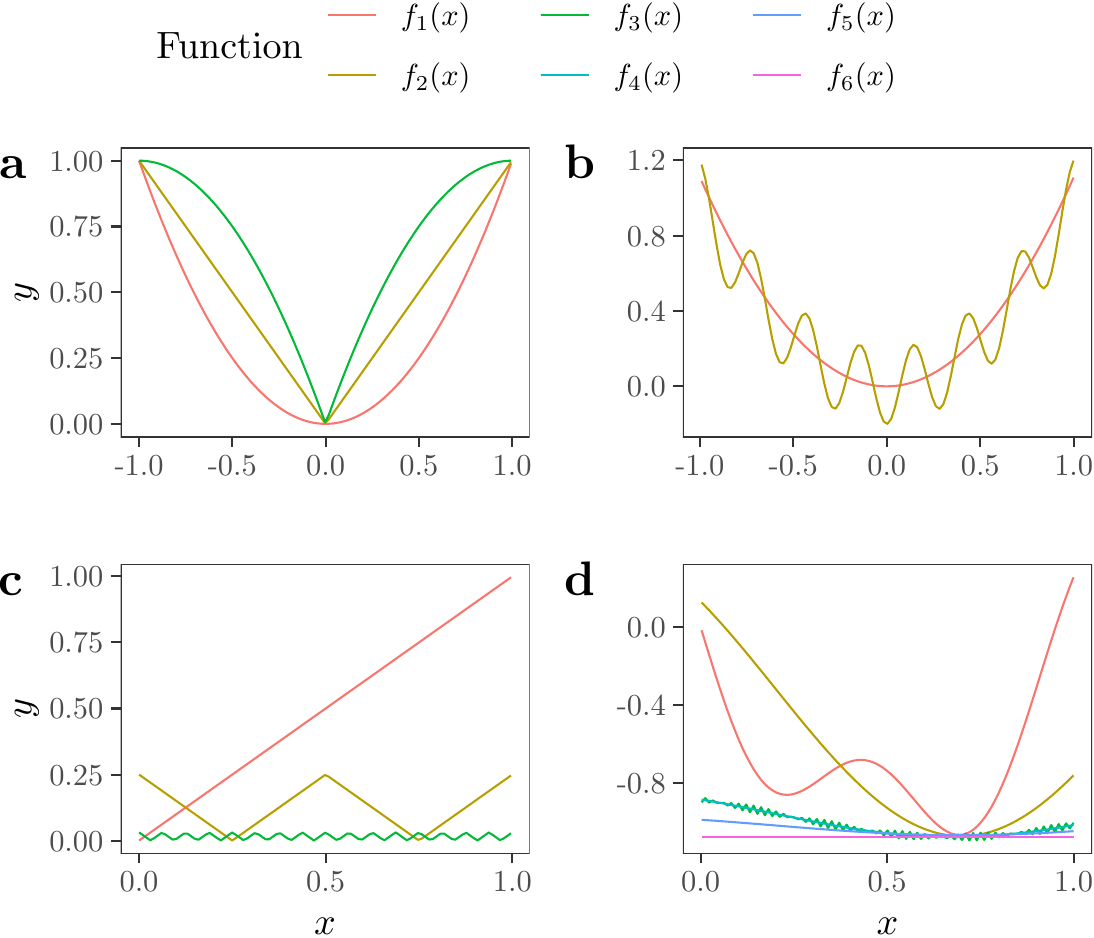}
\caption{Examples of functions in \textcite{Razavi2016a}. a) Unimodal functions with different structures. b) Multimodal versus unimodal function with identical variance. c) Functions covering different ranges in the response. d) A six-dimensional response surface. See the Supplementary Materials for a mathematical description of all functions in all sub-plots (see also Equation~\ref{eq:six_dimensional} for the model in d).}
\label{fig:first}
\end{figure}

\textcite{Razavi2016a} construct their case using several functions, which we reproduce hereafter. In Fig.~\ref{fig:first}a, Sobol' indices do not differentiate $f_3$ from $f_1$, whereas VARS points towards $f_3$ as the most sensitive function. In Fig.~\ref{fig:first}b, variance-based methods equate $f_1$ with $f_2$ because they have identical variance. According to \textcite[p.~428]{Razavi2016a}, this \say{runs counter to our intuitive notion of sensitivity} given the multimodality of $f_2$. If VARS is used, $f_2$ is identified as more sensitive than $f_1$ for $0\leq h\leq 0.2$. 

In Fig.~\ref{fig:first}c, Sobol' indices do not detect the periodicities of $f_2$, which Razavi \& Gupta suggest might be important in evaluating the impact of a factor from the perspective of model calibration. In Fig.~\ref{fig:first}d, variance-based methods regard $f_2$ as more sensitive than $f_1$. \textcite[p.~433]{Razavi2016a} argue that this is \say{contrary to intuition} because the effect of $f_1$ is more complex (bi-modal). IVARS$_{10}$ and IVARS$_{30}$, in contrast, characterise $f_1$ as more sensitive than $f_2$.

It is apparent that for \textcite{Razavi2016a} a sensitivity measure should be able to appraise the function structure. Our impression is that this perception of sensitivity is relevant to specific contexts, e.g. a diagnostic setting in which one is interested in the topology of a given function. However, the key lies in the definition of  \say{importance} pointed to by VARS. In which sense is $f_2$ more important than $f_1$ in Fig.~\ref{fig:first}b, or $f_3$ more important than $f_1$ and $f_2$ in Fig.~\ref{fig:first}a? If SA is used in an information quality setting \cite{Kenett2013}, when the aim is to determine which factor has the highest potential to reduce the uncertainty in the inference (i.e. how much is gained by discovering the true value of an uncertain factor), these functions might be regarded as equally sensitive. The same applies to Fig.~\ref{fig:first}d: given that $f_2$ changes more decidedly over the interval range than $f_1$, a larger reduction in uncertainty can be achieved by learning first about $f_2$ than about $f_1$.

Given that SA quantifies the relative influence of each model input in the model output, the concept of sensitivity is ultimately linked to that of \say{importance}. This is why it should be clear what do we mean when we say that a model input is \say{important}, or that a model output is very sensitive to a given model input. Variance-based methods meet this requirement by linking SA to statistical theory via ANOVA  \cite{Archer1997}, thus defining SA as \say{the study of how the variance in the model output is apportioned to different sources of uncertainty in the model input} \cite{Saltelli2004b}. The use of  variance-based methods such as Sobol' indices are well defined and associated with clear settings \cite{Saltelli2002c}:

\begin{enumerate}
\item Factors prioritization: the aim is to identify the single factor that, if determined (i.e., fixed to its true but unknown value), would lead to the greatest reduction in the variance of the output. This is met by the first-order sensitivity index ($S_i$).

\item Variance reduction: the aim is to identify the sets of factors (couples, triplets, and so on) leading to the reduction of the output variance below a given threshold, and doing this by fixing the smallest number of factors. This is achieved by using set (group) sensitivity indices.

\item Factors fixing: the objective is to identify factors that can be fixed anywhere in their range of variation without affecting the variance of the output. This is met by the total-order sensitivity index ($T_i$).
\end{enumerate}

Variance-based methods clearly resolve what is meant by \say{importance} of a factor. However, this is not as apparent in the case of VARS: if a decision needs to be taken based on the inference provided by a model, which of the variogram-based measures (IVARS$_{10}$, IVARS$_{30}$, VARS$_{50}$, VARS-TO) should be finally used to characterise the factors' importance? and what does \say{importance} mean for VARS? \textcite{Razavi2016a}'s statement of VARS being more \say{intuitive} than Sobol' indices is open to debate: intuition is in the eyes of the beholder, while solid criteria underpin the methodological quality of Sobol' indices.  

One way of gaining factual insight into the alleged ``intuitivity'' of VARS is through the analysis of its use by the 86 studies that have cited \textcite{Razavi2016a, Razavi2016b} up until November 2020. If adopted and used by practitioners other than the VARS authors themselves, and if the VARS framework is applied as recommended by their designers (i.e. by exploring different ranges of the spatial structure of the model response through integrated variograms IVARS and VARS-TO), the claim by \textcite{Razavi2016a, Razavi2016b} of VARS being an instinctive, user-friendly framework will find empirical support.

\begin{figure}[ht]
\centering
\includegraphics[keepaspectratio]{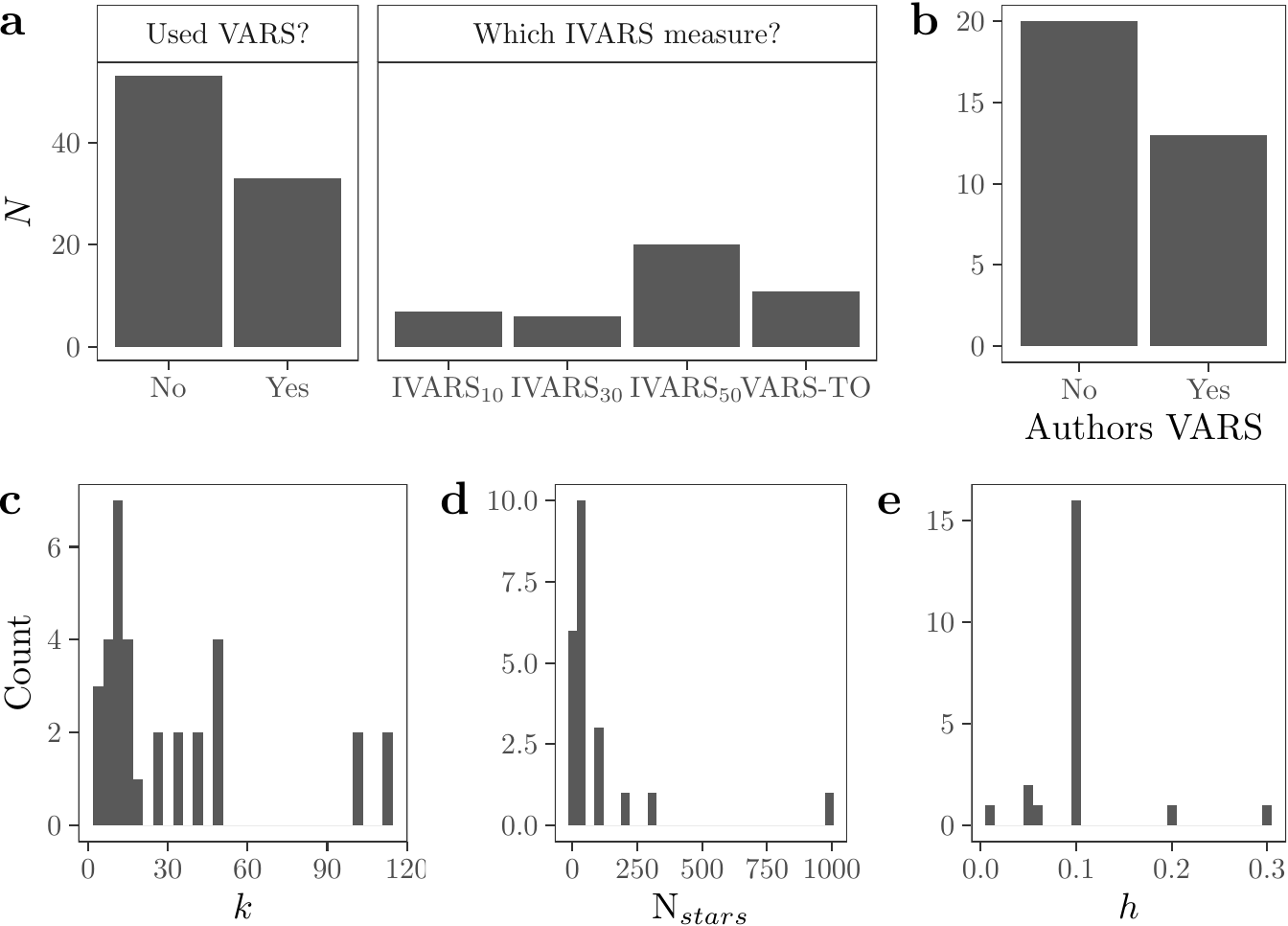}
\caption{Results of the survey conducted on all papers that cite \textcite{Razavi2016a, Razavi2016b} up to November 2020. a) Use of VARS.  b) Number of studies that use VARS and include (Yes) or do not include (No) the VARS authors themselves. c) Distribution of the dimensionality of the models for which VARS has been applied. d) Distribution of the number of stars used. e) Distribution of the values set for $h$.}
\label{fig:survey_results}
\end{figure}

We observed that 53 studies (62\%) cite \textcite{Razavi2016a, Razavi2016b} but do not implement VARS in any specific sensitivity analysis. Of the 33 studies that do apply VARS, 13 (40\%) include either Razavi and/or Gupta as lead author/s or co-authors. Hence the number of papers that use VARS and are not contributed by VARS authors amounts to 20, 23\% of all VARS citations (Fig.~\ref{fig:survey_results}a, b).

Out of the 33 studies that do use VARS, there were nine from which we could not retrieve precise information on the VARS metric/s used. As for the remaining 24, 15 studies used just one VARS metric (11 IVARS$_{50}$ and four VARS-TO), two used two (IVARS$_{50}$ and VARS-TO), three used three ([IVARS$_{10}$, IVARS$_{30}$, IVARS$_{50}$] x 2; IVARS$_{10}$, IVARS$_{50}$, VARS-TO) and four studies used all four metrics. The contributions by authors other than Razavi \& Gupta have strongly leaned towards the use of a single summary measure: out of the 12 works for which we could retrieve information on the VARS metric used, nine relied merely on one metric (six on IVARS$_{50}$ and three on VARS-TO), with one study using two, three and all four VARS measures.

With regard to the sensitivity settings, VARS has largely been applied to models with up to 20 parameters, with MESH being the model with the highest dimensionality (111). The number of stars has been mostly set between 20 and 50, with a single study raising it to 1000. A large number of works have used $h=0.1$, with the minimum and maximum $h$ values being 0.01 and 0.3 (Fig.~\ref{fig:survey_results}c--e).

As yet, such results place the ``intuitive nature'' of VARS in a disputable position: although cited, its use as a sensitivity measure has been comparatively discrete, and most authors have preferred a single summary VARS metric (IVARS$_{50}$ or VARS-TO, both very similar to the Sobol' total-order index \cite[p.~434]{Razavi2016a}) rather than implementing --and interpreting-- the whole integrated variogram approach.

The discussion above leads to another aspect listed by \textcite[p.~423]{Razavi2016a} as a motivation for developing VARS: an \say{ambiguous characterization of sensitivity}:

\begin{quotation}
\textit{(...) different SA methods are based in different philosophies and theoretical definitions of sensitivity. The absence of a unique definition for sensitivity can result in different, even conflicting, assessments of the underlying sensitivities for a given problem.}\footnote{The extent to which this points to an ambiguity is unclear. In any discipline, including statistics, different methods may naturally exist which become useful in different applications. For instance, the linear relation between two variables $x$ and $y$ might be modelled with Ordinary Least Squares (OLS) if $x$ causes $y$, or with Standard Major Axis (SMA) if it is unclear which variable is the predictor and which one is the response \cite{Smith2009a}. Does this mean that the characterisation of residuals in regression analysis is an ambiguous branch of statistics?}
\end{quotation}

We argue that the source of ambiguity in sensitivity analysis is not the lack of a unifying theory, or the fact that many sensitivity measures are available, but in the definition of \say{importance}. Unless the analyst stipulates what she means when she says that a variable is important, different methods can be thrown at the model resulting in different ordering of importance of the input variables, whereby the analyst could be tempted to cherry-pick the method most conforming to one's own bias. By linking the definition of importance to clear settings, Sobol’ indices resolve this quandary clearly and transparently, and provide end-users with a plain English description of the results. This comes in handy when the receiver (customer) of the analysis is not another practitioner. 

\begin{figure}[ht]
\centering
\includegraphics[keepaspectratio]{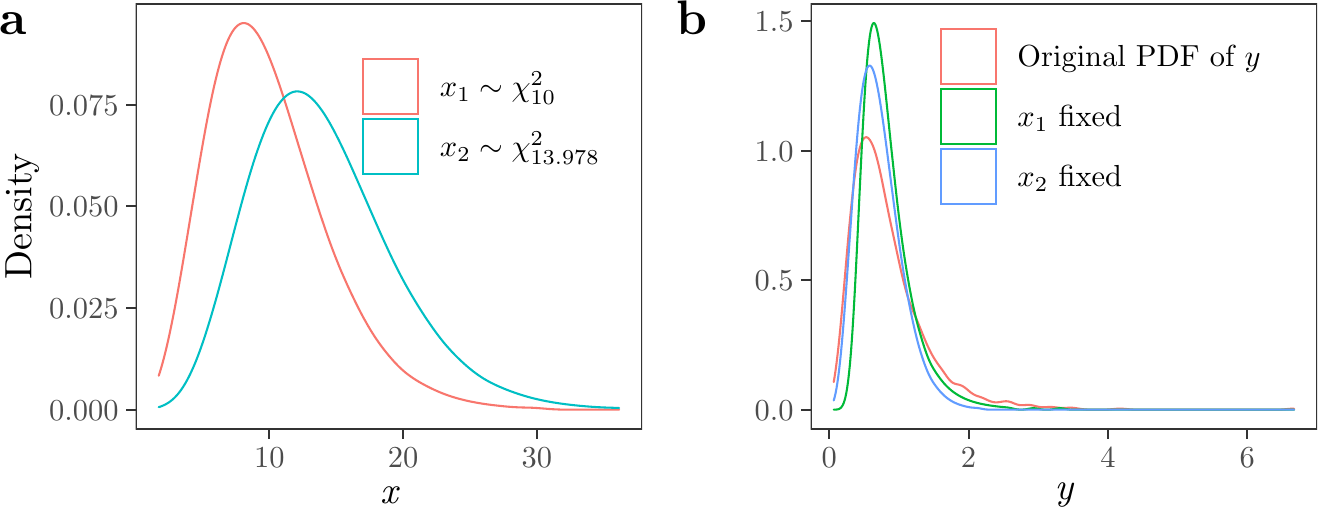}
\caption{The highly skewed function of \textcite{Liu2006a}. a) Distribution of $x_1$ and $x_2$. b) Comparison of impacts of inputs.}
\label{fig:liu}
\end{figure}

The expedient to produce functions where Sobol' indices look \say{wrong} is quite common. This approach was also taken by \textcite{Liu2006a} and \textcite{Pianosi2015} with Liu's highly-skewed function $y=\frac{x_1}{x_2}$, where $x_1\sim \chi^2(10)$ and $x_2\sim \chi^2(13.978)$ (Fig.~\ref{fig:liu}). The reader might wonder why one of the degrees of freedom is expressed with two-digit precision and the other with a five-digits one. The reason is that, with these crisp numbers, $T_1$ and $T_2$ are identical and equal to $0.5462$, while inspection of Fig.~\ref{fig:liu}b should convince us that $x_1$ is more important than $x_2$ by virtue of its longer tail. The Liu function is thus what \textcite{Lakatos1976} would have called a monster example, designed on purpose to invalidate variance-based methods. However, based on the definition of \say{importance} of Sobol' indices, the fact that they are equally influential appears totally reasonable.

We conclude by stating that rather than hinting at what should or not should be intuitive, a sensitivity index should pin down its definition of importance in unambiguous terms.

\section{The issue of efficiency}
\textcite{Razavi2016a, Razavi2016b} claim that VARS-TO is much more computationally efficient than the total-order estimator of \textcite[Eq.~4.23]{Saltelli2008} (up to two orders of magnitude), which is taken as a state-of-the-art implementation of the Sobol' approach. They make their case with three different models:

\begin{enumerate}
\item The six-dimensional response surface displayed in Fig.~\ref{fig:first}d, which is a purely additive model. VARS-TO accurately ranks the model inputs with just 60 simulations, beating the \textcite{Saltelli2008} estimator of total-order indices at $>6,000$ simulations \cite[pp.~435--436]{Razavi2016a}.

\item The five-dimensional conceptual rainfall-runoff model HYMOD \cite{Vrugt2003}. VARS-TO detects the \say{true} ranking of the model inputs at 500 simulations, while the \textcite{Saltelli2008} estimator requires 10,000 simulations \cite[pp.~443--444]{Razavi2016b}.

\item The 45-dimensional land surface scheme-hydrology model MESH \cite{Pietroniro2007}. The VARS-TO estimate of the total-order effect stabilizes at 5,000 simulations, whereas the \textcite{Saltelli2008} estimator requires more than 100,000 simulations \cite[pp.~453--454]{Razavi2016b}.
\end{enumerate} 

Do these examples truly prove that VARS-TO is between 20 and 100 times more efficient than the Sobol'-based approach to total-order indices? 

\subsection{The case of the six-dimensional response surface model}
\label{sec:six_dimensional}
To properly answer this question in the case of the six-dimensional model (Fig.~\ref{fig:first}d), whose functional form reads as
\begin{equation}
\begin{aligned}
g_1(x_1) & =-\sin (\pi x_1) - 0.3 \sin (3.33\pi x_1) \\
g_2(x_2) & = -0.76 \sin \left [ \pi (x_2 - 0.2) \right ] - 0.315 \\
g_3(x_3) & = -0.12 \sin \left [1.05\pi(x_3 - 0.2) \right ] - 0.02 \sin (95.24 \pi x_3)-0.96 \\
g_4(x_4) & = -0.12 \sin \left [ 1.05 \pi(x_4 - 0.2)\right] - 0.96 \\
g_5(x_5) & = -0.05 \sin \left[ \pi ( x_5 - 0.2) \right ] - 1.02 \\
g_6(x_6) & = -1.08 \\
y & =f \left [ g_1(x_1) + g_2(x_2) +...+g_6(x_6) \right ]\,,
\end{aligned}
\label{eq:six_dimensional}
\end{equation}
we should first focus on the sampling design of VARS and Sobol' indices. 

\begin{figure}
\centering
\begin{minipage}{0.47\textwidth}
\centering
\includegraphics[width=0.9\textwidth]{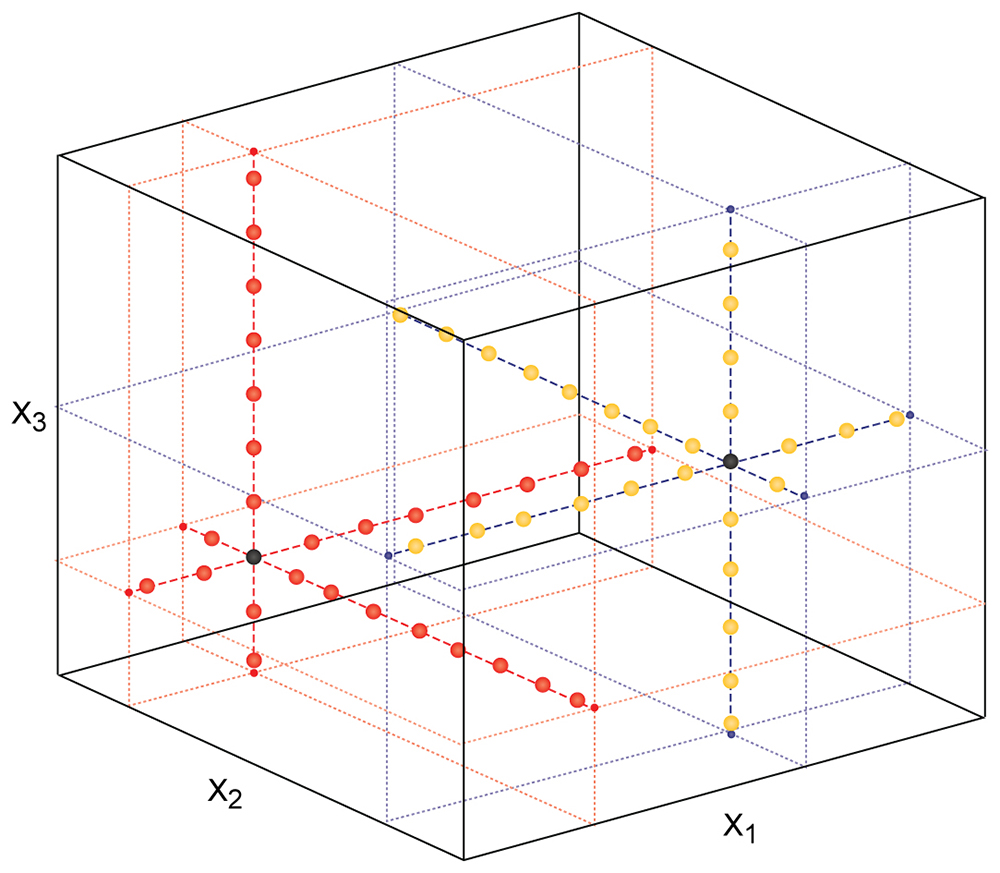} %
\end{minipage}\hfill
\begin{minipage}{0.47\textwidth}
\centering
\includegraphics[width=1\textwidth]{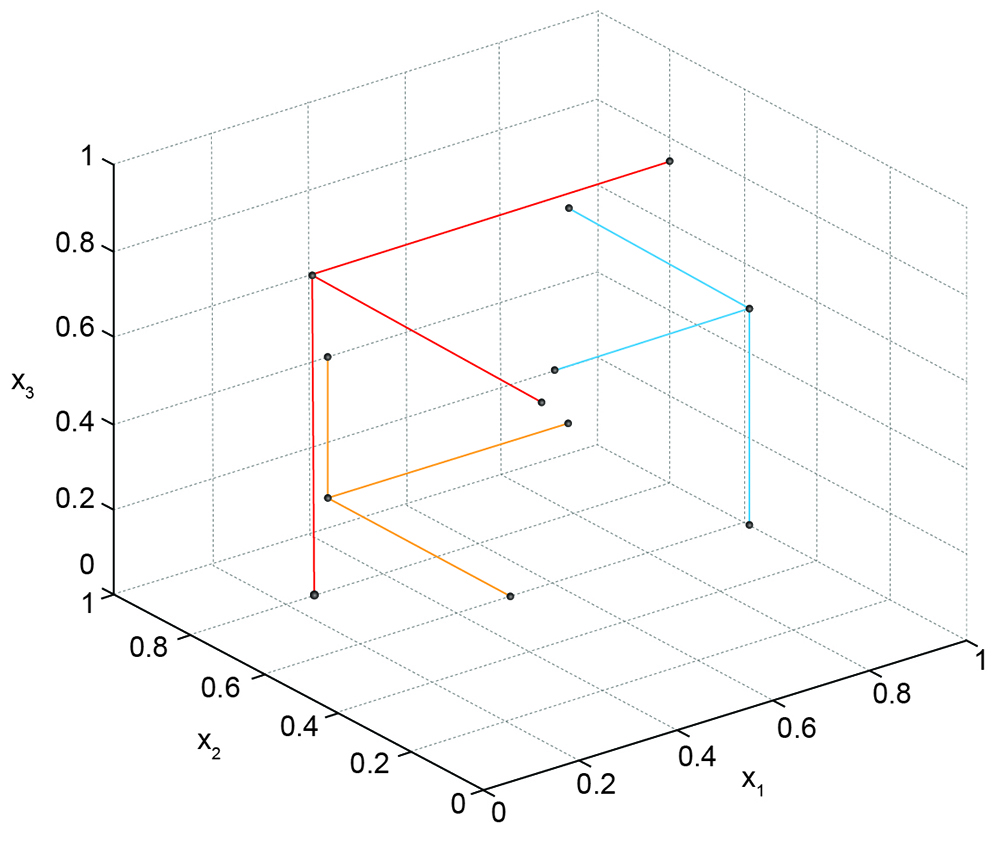} %
\end{minipage}
 \caption{Sampling design of VARS-TO (STAR-VARS, left) and Sobol' indices (right). For VARS-TO the plot shows a star-based sampling in three dimensions, with $\Delta h=0.1$ and number of stars $N_{star}=2$. Black dots are the star centers, and the coloured dots are the additional $\frac{1}{\Delta h}$ points along the three axes. Adapted from \textcite[Fig.~1]{Razavi2016b}. For Sobol' based-indices, the plot also displays a three-dimensional model, with the links being steps in the $x_i$ direction and $N=4$. Adapted from \textcite[Fig.~18.7]{Becker2014}.}
\label{fig:star}
\end{figure}

The computation of VARS relies on stars and is referred to as STAR-VARS by \textcite{Razavi2016b}: the analyst first randomly selects $N_{star}$ points across the factor space, i.e. via random numbers, Latin Hypercube Sampling (LHS) or Sobol' Quasi Random Numbers (QRN). These are the \say{star centres} and their location can be denoted as $\bm{s}_v = s_{v_1},...,s_{v_i}, ..., s_{v_k}$, where $v=1,2,...,N_{star}$. Then, for each star centre, a cross section of equally spaced points $\Delta h$ apart needs to be generated for each of the $k$ factors, including and passing through the star centre (Fig.~\ref{fig:star}, left side plot). The cross section is produced by fixing $\bm{s}_{v_{\sim i}}$ and varying $s_i$. Finally, for each factor all pairs of points with $h$ values of $\Delta h, 2\Delta h, 3\Delta h$ and so on should be extracted. The total computational cost of this design is $N_t=N_{star} \left [ k (\frac{1}{\Delta h} - 1) + 1 \right ]$.

Sobol' indices also rely on a star-based sampling strategy: they require a $(N, 2k)$ base sample matrix, designed via LHS or QRN, in which the rightmost $k$ columns are allocated to an $\bm{A}$ matrix and the leftmost $k$ columns to a $\bm{B}$ matrix. Then, $k$ extra $(N, k) \bm{A}^{(i)}_B$ matrices are created, where all columns come from $\bm{A}$ except the $i$-th, which comes from $\bm{B}$. This design creates stars with centres and points a step away in the $x_i$ direction (Fig.~\ref{fig:star}, right-side plot). The cost of this design for $T_i$ is $N_t=N(k+1)$, where $N$ is the row dimension of the base sample matrix.

When the function or the model under study is fully additive, as in the six-dimensional surface model mentioned above (Fig.~\ref{fig:first}d), the computation of VARS-TO can be done with a single cross-section in the space of $x^*_{\sim i}$ for each model input. VARS-TO thus becomes a first-order index \textit{de facto}, as one model input remains constant while all the others vary. The natural term of comparison is thus the Sobol' first-order index, and not the total. In that sense, and for any function which behaves non-additively for at least one factor, i.e. $F=f(x_i) + g(\bm{x}_{\sim i})$, the first-order effect $S_i$ can be computed very easily, since 
\begin{equation}
S_i=\frac{E_{x_i} \left [ f(x_i) \right ] ^2-E_{x_i}^2 \left [ f(x_i) \right ] }{V(F)}\,,
\label{eq:one_trajectory}
\end{equation}
i.e. $S_i$ is only a function of $x_i$ and hence it can be computed with a single trajectory along $x_i$, irrespective of its position in $\bm{x}_{\sim i}$. We provide the proof in Section 2.1 of the Supplementary Materials. 

\begin{figure}[ht]
\centering
\includegraphics[keepaspectratio]{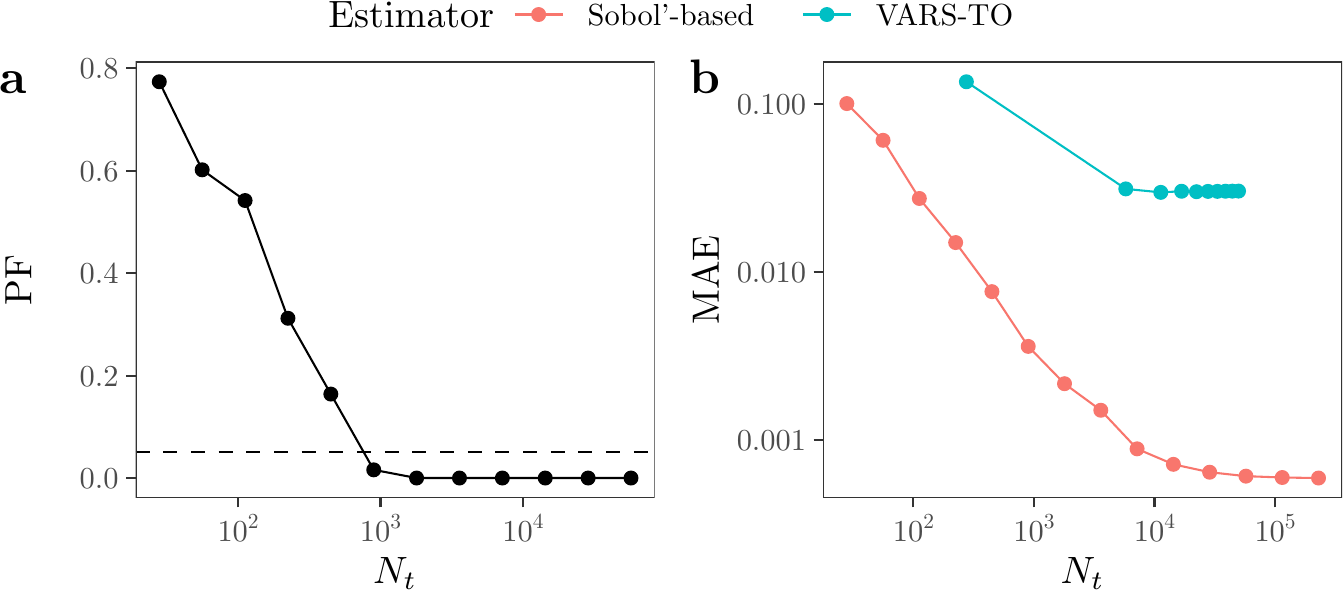}
\caption{Assessment of VARS-TO and the Sobol'-based estimator (Equation~\ref{eq:one_trajectory}) with the six-dimensional model as a test function. a) Probability of failure (PF) of Equation~\ref{eq:one_trajectory} in correctly ranking the model inputs. Each dot summarises the PF over 500 quasi-random number matrices with different starting points. The horizontal dashed line is at PF=0.05. For more details about the computational methodology, see \textcite{Razavi2016a}. b) Mean Absolute Error (MAE) (see Equation~\ref{eq:MAE}), with $p=50$. See Section 2.1.1 of the Supplementary materials for the computation of the analytical values of the six-dimensional model.}
\label{fig:PF}
\end{figure}

We used Equation~\ref{eq:one_trajectory} to compute $S_i$ for the six-dimensional model, aiming at replicating the results by \textcite[see their Fig.~6]{Razavi2016a}. For VARS-TO and Sobol'-based indices they tested their probability of failure, defined as the probability of obtaining erroneous ranks for the model inputs of the six-dimensional model (Fig.~\ref{fig:first}d, Equation~\ref{eq:six_dimensional}). We observed that, if Equation~\ref{eq:one_trajectory} is used to compute Sobol'-based indices, all model inputs are accurately ranked at $N_t=896$ (Fig.~\ref{fig:PF}a), contrasting with the $N_t > 6,000$ obtained by \textcite{Razavi2016a}. This example suggests that VARS-TO is indeed more efficient than a Sobol'-based approach when the model is fully additive and the aim is to rank the parameters, but significantly less than what the authors claimed it to be. It is also worth noting  that there are other approaches that might permit a more efficient computation of first-order indices \cite{Plischke2010, Mara2017, Strong2012}.

However, why do Razavi \& Gupta rely exclusively on the ``probability of failure'' in the ranking as a performance metric? Sorting the parameters by their influence in the model output is indeed a common setting in sensitivity analysis. But there are other goals: the analyst might be more interested in getting exact values for the sensitivity indices in order to ascertain, for instance, how much the uncertainty would be reduced if the ``true'' value of an uncertain factor is discovered. In such context, a performance measure such as the mean absolute error (MAE) between the estimated ($\bm{\hat{T}})$ and the analytical ($\bm{T}$) values might be more appropriate. The MAE has been a very widespread performance measure in sensitivity analysis \cite{LoPiano2020, Saltelli2010a}, and is computed as
\begin{equation}
\text{MAE}=\frac{1}{p}\sum_{v=1}^{p} \left ( \frac{\sum_{i=1}^{k} | T_i - \hat{T}_i|}{k} \right )\,,
\label{eq:MAE}
\end{equation}
where $p$ is the number of replicas of the sample matrix,  and $T_i$ and $\hat{T}_i$ the analytical and the estimated total-order index of the $i$-th input.

Had Razavi \& Gupta relied on the MAE rather than on the ``probability of failure'' as a performance measure, their assessment of the efficiency of VARS-TO and Sobol'-based indices would have been very different: throughout the range of explored model runs, VARS-TO never approaches Equation~\ref{eq:one_trajectory} in terms of accuracy (Fig.~\ref{fig:PF}b). These results exemplify how sensitive the results of a benchmarking exercise can be to the particular settings defined by the analyst: merely the use of a different performance measure can completely tip the balance from one estimator to another. In the section below, we show how to minimize this source of bias to get a more accurate picture of the true performance of VARS-TO compared to Sobol'-based estimators \cite{Puyj}.

\subsection{The case of the HYMOD and MESH models}

Unlike the six-dimensional model, HYMOD and MESH are non-additive models. Hence a single trajectory is not enough and several cross-sections in the space of $x^*_{\sim i}$ should be drawn to fully explore the hypercube. Under such settings, the comparison between VARS-TO and a Sobol'-based estimator of the total-order index is the appropriate methodological choice.

But does the higher accuracy of VARS-TO reported by \textcite{Razavi2016a, Razavi2016b} for these two models truly evidence its superiority over Sobol'-based total-order indices? We argue that the following issues make \textcite{Razavi2016a, Razavi2016b}'s claim controversial:

\begin{itemize}
\item The use of the \textcite{Saltelli2008} total-order estimator as ``state-of-the-art''. Amongst all estimators available for computing $T_i$, that of \textcite{Saltelli2008} ranks close to last on accuracy and performance and is significantly outperformed by the Jansen or the Janon / Monod estimators \cite{Jansen1999, Monod2006a, Janon2014, Puyj}. Furthermore, \textcite{Saltelli2010a} demonstrated that configurations based on $\bm{B}$, $\bm{B}^{(i)}_A$ matrices (as is the case of the Saltelli  estimator) were surpassed in performance by those relying on $\bm{A}$, $\bm{A}^{(i)}_B$ matrices (i.e. the Jansen estimator) when Quasi-Random numbers were used to create the sample matrix.

\item The extrapolation of the results obtained with HYMOD and MESH to mean that VARS-TO is \emph{generally} better than Sobol'-based indices. \textcite{Puyj} recently showed that, once the benchmark settings are randomised (i.e. the model and its dimensionality, the sampling method, the total number of model runs, the fraction of active second and third-order effects, the distribution of the model inputs and the performance measure), VARS-TO loses much of its purported computational superiority: it only very slightly outperforms the Sobol'-based estimators \textcite{Jansen1999} and Janon/Monod \cite{Janon2014, Monod2006a} when there are serious constraints on the number of model runs that can be allocated to each model input (i.e. 2-10). At larger sample sizes, the performances of VARS and Jansen and Janon/Monod are exactly the same \cite{Puyj}. This randomisation is required to reduce the dependence of the results on the choices taken by the analyst: as we have seen in the case of the six-dimensional model (see Section~\ref{sec:six_dimensional}), even the use of a given performance measure over another might completely swap the outcome of an analysis.
\end{itemize}

In order to obtain a more comprehensive view of the performance of VARS-TO against Sobol' indices, we have reproduced the work by \textcite{Puyj} with the following changes and/or additions:

\begin{enumerate}
\item We have tested VARS-TO against the \textcite{Jansen1999} formula, one of the most precise and accurate Sobol' total-order estimators \cite{Puyj}. 

\item We have increased the range of the proportion of active second and third-order effects in the test functions (i.e. between 50-100\% and 30-100\% respectively; in \textcite{Puyj} they ranged between 30-50\% and 10-30\% respectively ). This aimed at checking how VARS-TO performs under serious non-additivities.

\item We have taken into account the algorithmic uncertainties of VARS-TO, i.e. the number of stars $N_{star}$ and the distance between pairs $h$, which ultimately condition its computational cost. These design parameters need to be set by the analyst before executing the algorithm, and the specific value that might work best is unclear. Different authors have used different values for $h$ ($\Delta h = 0.002, \Delta h = 0.1, \Delta h = 0.3$; \cite{Becker2019, Razavi2016a, Razavi2016b, Krogh2017}; see Fig.~\ref{fig:survey_results}e). \textcite{Razavi2019a} recommend $h=0.1$ and $h<0.1$ if more accurate results are needed. As shown by \textcite{Puy2020} for PAWN, the uncertainty in the design parameters of a sensitivity index might contribute appreciably to its volatility.
\end{enumerate}

We compared the performance of VARS-TO and the Jansen estimator by treating the main benchmark settings listed in Table~\ref{tab:parameters} as uncertain parameters described by probability distributions. We created a $(2^{12}, k)$ sample matrix using Sobol' quasi-random numbers \cite{Sobol1967, Sobol1976}, in which each row is a sample point and each column an uncertain parameter. For $v=1,2,...,2^{12}$ rows, we computed VARS-TO and the Jansen total-order index according to the specifications set by $N_{{star}_v}, h_v,...,\delta_v$. The final model output was $r_v$, the correlation coefficient between the indices estimated by VARS-TO and Jansen ($\bm{\hat{T}}_v$) and the \say{true} indices ($\bm{T}_v$), computed with a large sample size ($N=2^{12}$) and the \textcite{Jansen1999} estimator. The larger the $r_v$, the better the estimation of the "true" sensitivity indices by VARS-TO or Jansen. We argue that this approach allows us to examine the accuracy of VARS-TO more comprehensively given the enormous range of sensitivity problems that it is able to explore \cite{Becker2019, Puyj} (potentially more than 3 billion scenarios in this case). If VARS-TO beats Sobol'-based indices unequivocally, as asserted by \textcite{Razavi2016a, Razavi2016b}, its computational advantage should emerge against Jansen as well. The Supplementary Materials thoroughly detail the rationale and the execution of the experiment.

\begin{table}[ht]
\centering
\caption{Summary of the uncertain parameters and their distributions. $\mathcal{DU}$ is discrete univariate. See the Supplementary Materials for a description of the rationale behind the selection of the uncertain parameters and their distributions.}
\label{tab:parameters}
\begin{tabular}{llc}
\toprule
Parameter & Description & Distribution \\
\midrule
$N_{star}$ & Number of star centres & $\mathcal{DU}(3, 50)$ \\
$h$ & Distance between pairs & $\mathcal{DU}(0.01, 0.05, 0.1, 0.2)$\\
$k$ & Number of model inputs & $\mathcal{DU}(3,50)$ \\
$\varepsilon$ & Randomness in the test function & $\mathcal{DU}(1, 200)$ \\
$\tau$ & Sampling method & $\mathcal{DU}(1,2)$ \\
$\phi$ & Probability distribution of the model inputs & $\mathcal{DU}(1, 8)$ \\
$k_2$ & Fraction of pairwise interactions & $\mathcal{DU}(0.5, 1)$ \\
$k_3$ & Fraction of three-wise interactions & $\mathcal{DU}(0.3, 1)$ \\
$\delta$ & Selection of the performance measure & $\mathcal{DU}(1, 3)$ \\
\bottomrule
\end{tabular}
\end{table}

\begin{figure}[!ht]
\centering
\includegraphics[keepaspectratio]{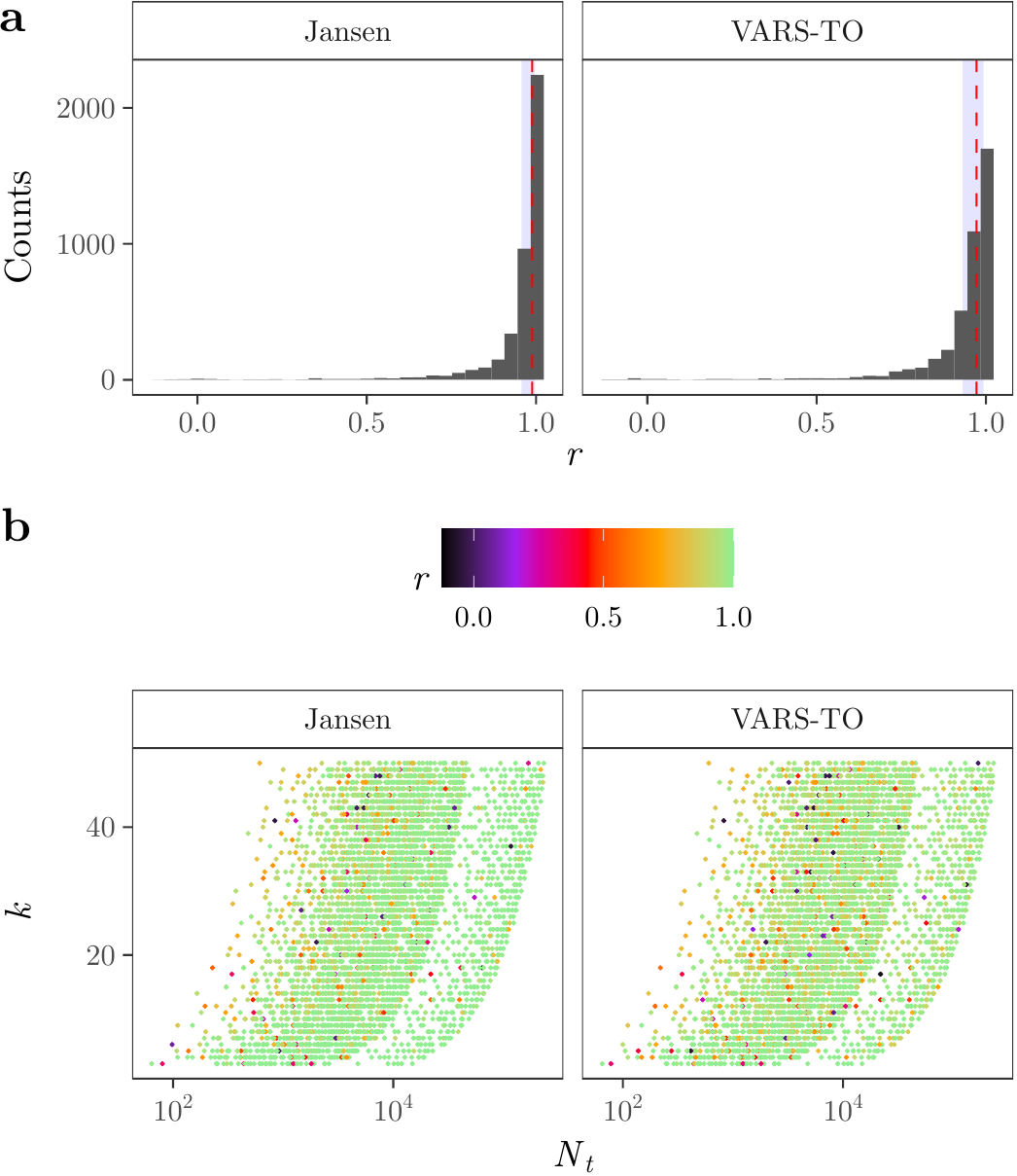}
\caption{Uncertainty analysis conducted on $2^{12}$ simulations. a) Histograms of $r$. The vertical red, dashed line shows the median value. The transparent, blue rectangle frames the 0.25, 0.75 quantiles. b) Scatterplots showing the performance of Jansen and VARS-TO as a function of the total number of model runs $N_t$ and the model dimensionality $k$. Each simulation is a dot. The greener (darker) the colour, the better (worse) the performance. The white space between $10^3$ and $10^4$ in the $x$ axis is caused by the uneven distribution selected for $h$ (see Table~\ref{tab:parameters} and the Supplementary Materials).}
\label{fig:uncertainty}
\end{figure}

\begin{figure}[!ht]
\centering
\includegraphics[keepaspectratio]{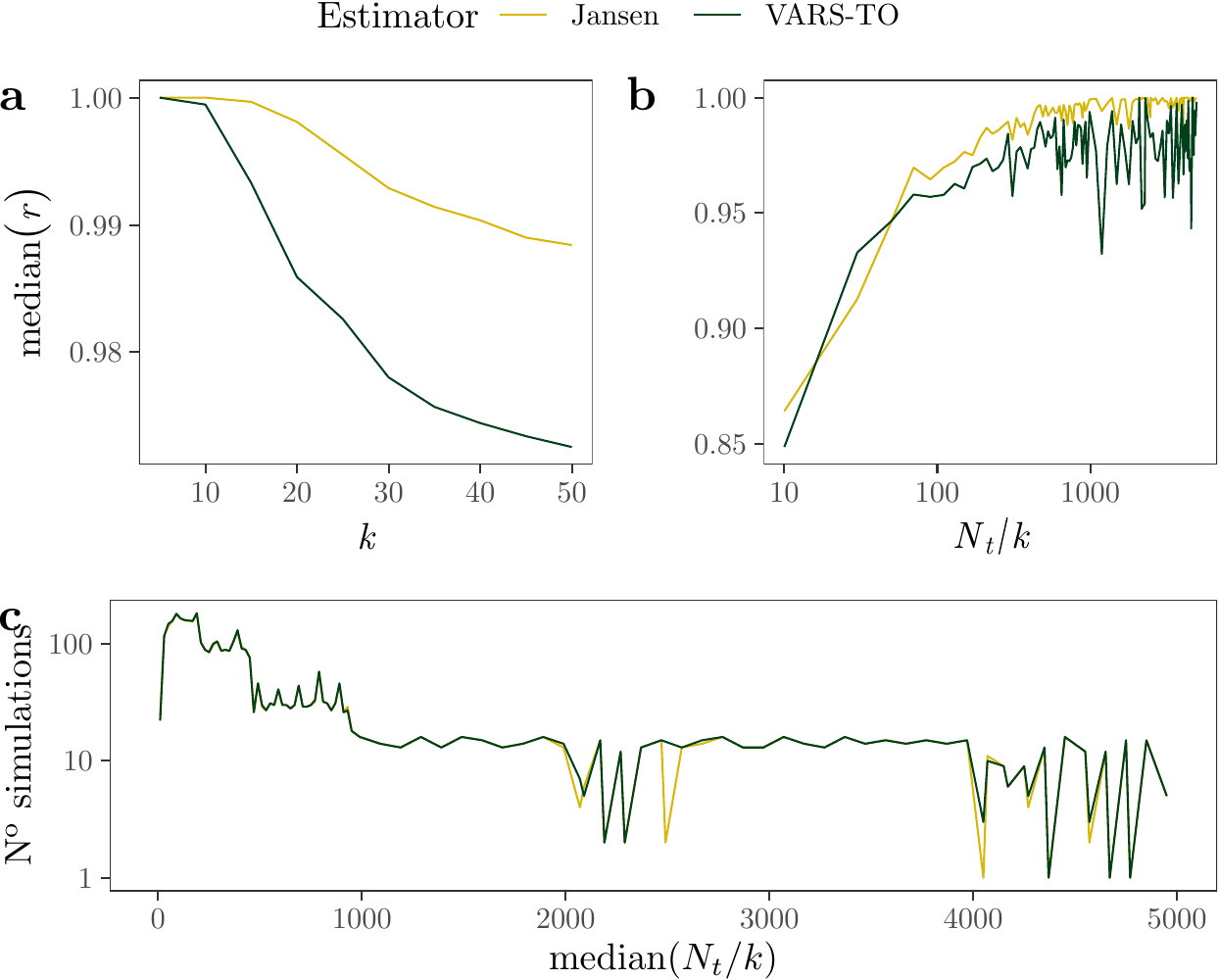}
\caption{Comparison between the accuracy and efficiency of VARS-TO and \textcite{Jansen1999}. a) Evolution of the median $r$ value across different dimensions $k$. b) Evolution of the median $r$ value across the different number of runs allocated to each model input ($N_t / k$). c) Number of simulations with the same $N_t / k$ ratio. Both lines almost fully overlap.}
\label{fig:medians}
\end{figure}

Fig.~\ref{fig:uncertainty}a shows that both Jansen and VARS-TO are very accurate as the empirical distribution of $r$ is highly right-skewed. If anything, Jansen seems to outperform VARS-TO overall due to its slightly narrower distribution (95\% CI 0.93-0.99, median of 0.99 for Jansen; 95\% CI 0.87-0.99, median of 0.97 for VARS). This is also apparent in Fig.~\ref{fig:uncertainty}b, with VARS-TO presenting more simulations with redder/orange colours (approx. $r \leq 0.85$). A closer look at the performance of both approaches reveals that Jansen maintains a higher median accuracy at higher dimensions (Fig.~\ref{fig:medians}a), whereas VARS-TO confirms its slightly higher efficiency only when the number of runs that can be allocated per model input ($N_t / k$) is considerably constrained ($<50$ in this case, Fig.~\ref{fig:medians}b) \cite{Puyj}. VARS-TO also displays a larger volatility at $100>(N_t / k)$ (Fig.~\ref{fig:medians}b), suggesting that Jansen might become more stable in a larger number of sensitivity problems if the number of model runs per input is increased. These results rest on solid grounds as the number of simulations for which we have computed the median $N_t / k$ is almost identical for Jansen and VARS-TO (Fig.~\ref{fig:medians}c). Overall, this proves that both estimators have a very similar efficiency and reliability.

\begin{figure}[ht]
\centering
\includegraphics[keepaspectratio]{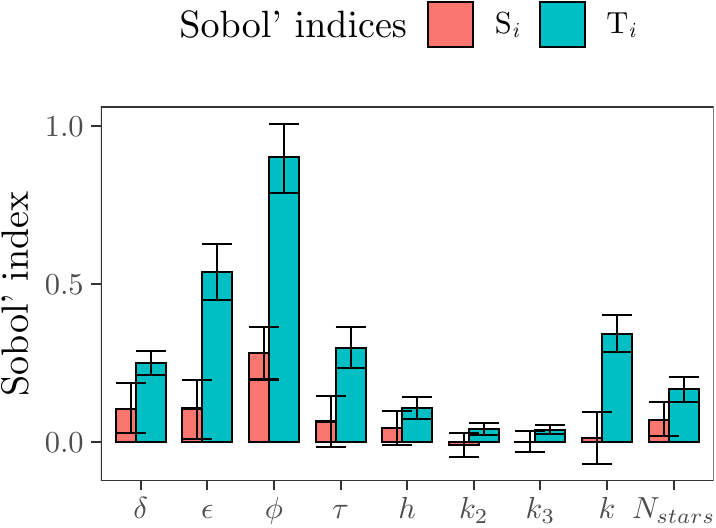}
\caption{Sobol' indices for VARS-TO. The error bars show the 95\% confidence intervals, computed with the percentile method after bootstrapping ($R=500$). }
\label{fig:sobol}
\end{figure}

We also computed Sobol' indices to assess which uncertain parameter most influences the performance of VARS-TO (Fig.~\ref{fig:sobol}). We observed that c. 30\% the variance in its performance is driven by the underlying probability distribution of the model inputs $\phi$, which appears as the most influential parameter. The other parameters are important through interactions, especially the functional form of the model ($\epsilon$), the sampling method ($\tau$), the model dimensionality ($k$) and the performance measure selected ($\delta$), in that order. The proportion of second and third-order effects ($k_2,k_3$) has no effect, which means that VARS-TO is very robust against non-additivities.

\begin{figure}[!ht]
\centering
\includegraphics[keepaspectratio]{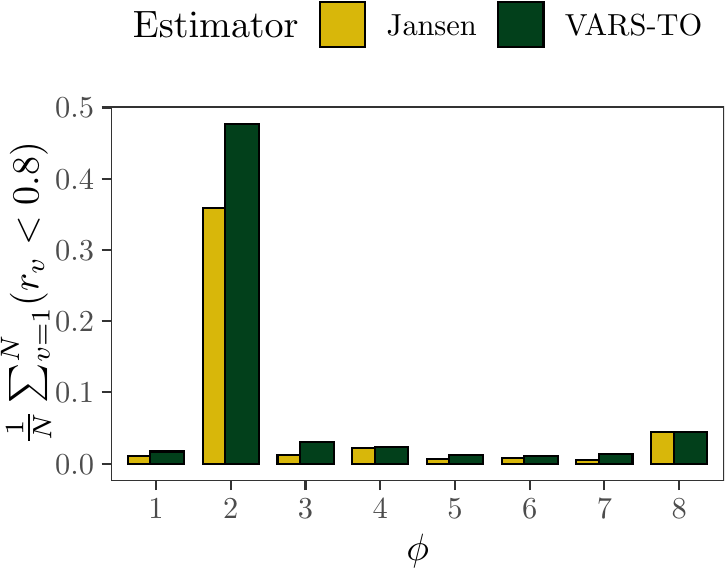}
\caption{Proportion of model runs with $r < 0.8$ as a function of $\phi$. The normal distribution $\left [ \mathcal{N}\sim(0.5, 0.2) \right ]$ is triggered by $\phi=2$. See Fig.~S1 for a recoding of all levels of $\phi$ into their probability distributions.}
\label{fig:normal}
\end{figure}

Compared to Jansen, VARS-TO significantly underperformed when the model inputs were normally distributed (e.g. when $\phi=2$, Figs.~S5, \ref{fig:normal}). We observed that this was caused by high-order interactions between the sampling design of VARS-TO (Fig.~\ref{fig:star}, left side) and at least five different uncertain parameters, $N_{star},h,k,\phi,\tau$. 

To understand these interactions, let us first assume that we use random numbers ($\tau=1$) to sample our star centers, which \textcite[Table~1]{Razavi2019a} list as a possible sampling strategy to compute VARS-TO. These star centers are located at $\bm{s}_v = s_{v_1},...,s_{v_i}, ..., s_{v_k}$, where $v=1,2,...,N_{star}$. The higher the $N_{star}$ and $k$, the higher the chances that a value at the boundary of $(0,1)$ is included in $\bm{s}_v$. Given that VARS-TO requires fixing $\bm{s}_{v_{\sim i}}$ while varying $s_i$ at a step defined by $h$, this value at the periphery of $(0,1)$ will be repeated in the $ \left [ (\frac{1}{h}) - 1 \right ] (k-1)$ coordinates, which can be manifold if $k$ is high and $h$ is low. Once the model inputs are transformed into a normal distribution, it will turn into an extreme value and will disrupt both the model output and the computation of VARS-TO for the $\bm{x}_{\sim i}$ parameters.

Let us now assume that we do not use random numbers to sample the star centres, but Sobol' Quasi-Random (QRN) number sequences ($\tau=2$). They are also contemplated by \textcite[Table~1]{Razavi2019a} as a sampling strategy to compute VARS-TO. Although the design of QRN makes the sampling of star centres at the very periphery $(0,1)$ very unlikely, cross-sections can indeed sample the boundary of the domain: for instance, if $h=0.1$ and $s_{v_i}=0.5$, the cross-section of the $x_i$ parameter will be the vector $\bm{x}_i=0.1, 0.2,... s_{v_i}, ..., 1$. This will will cause VARS-TO to crash as a uniform one becomes infinity under a normal distribution. Even if the STAR-VARS algorithm is modified to prevent 1 from being sampled (e.g. by replacing 1 by 0.999, as we did in this study), some cross-sections will still sample values very close to 1 by design, especially if $h$ is set at a small value. These values will be extreme values under a normal distribution, disrupting again the computation of the model output and the VARS-TO index --in this case, for the $x_i$ parameter.

We believe that this explains the high-order interactions involving $N_{star},h,k,\phi,\tau$, which are non-negligible (Fig.~\ref{fig:sobol}). The effect of $N_{star}$ and $h$ in VARS-TO was not explored by \textcite{Puyj} nor by \textcite{Becker2019}, who documented a slightly higher performance of VARS-TO against Jansen and Janon/Monod. Our results indicate that VARS-TO loses this marginal edge once these internal uncertainties jointly with the uncertainties in the benchmark settings are considered in the computations. Even if the use of VARS-TO is restricted to non-normal distributions ($\phi \neq 2$), its performance would still be slightly outdone by Jansen (95\% CI 0.96-0.99, median of 0.99 for Jansen; 95\% CI 0.94-0.99, median of 0.97 for VARS).

\section{Conclusions}
We have revised the Variogram Analysis of Response Surfaces (VARS), a new framework for sensitivity analysis developed by \textcite{Razavi2016a, Razavi2016b}. We have specifically focused on two aspects that, according to \textcite{Razavi2016a, Razavi2016b}, make VARS outperform Sobol' indices: its more intuitive appraisal of sensitivities and of the importance of model inputs, and its 20-100 times higher computational efficiency.

The claim that VARS is more intuitive than Sobol' indices can hardly be reversed as it ultimately is a matter of personal taste, disciplinary orientation and objective of the modeling activity: a geographer working in a diagnostic model setting might indeed prefer VARS's approach to the model structure. However, the professed higher intuitive nature of VARS ties in poorly with the available evidence: less than one half of the studies citing VARS actually implement it in a case study, and almost half of those that use it include the VARS authors themselves. Furthermore, most works do not explore the full range of the response surface but rely exclusively on summary metrics such as IVARS$_{50}$ or VARS-TO, which are very similar to the Sobol' total-order index. 

We argue that Sobol' indices provide a clearer description of what an \say{important} model input is given its connection to statistical theory and ANOVA. The use of Sobol' first or total order indices is associated with clear research settings and their meaning can be easily conveyed to stakeholders or non-specialists, which adds to their transparency. VARS, in contrast, allows the analyst to zoom into the structure of the model output and assess its dependency on the model inputs through the integrated variograms IVARS$_{10}$, IVARS$_{30}$ and IVARS$_{50}$, as well as through the variance-based total-order effect VARS-TO. But what is their definition of importance? how useful is it for a stakeholder to know that a parameter is \say{important} under IVARS$_{10}$ and not as much under IVARS$_{30}$, for instance? Which IVARS measure should she finally rely onto before making a policy decision? If the answer is the summary measure VARS-TO, then it is unclear how VARS advances Sobol' indices given the reliance of VARS-TO on variance and covariance matrices.

The purported much higher efficiency of VARS-TO is very contentious. The observation that it is more than 100 times more efficient than Sobol'-based total-order indices rests on an exercise with a fully additive model, in which VARS-TO is compared against one of the less accurate total-order estimators (\textcite{Saltelli2008}, see \textcite{Puyj}), and the performance measure chosen is the ``probability of failure'' of properly ranking the model inputs. If the comparison is instead conducted between VARS-TO and a lightweight Sobol'-based first-order estimator (Equation~\ref{eq:one_trajectory}), the advantage of the former shrinks to being ``only'' 15 times more efficient. VARS-TO completely loses all its edge if the Mean Absolute Error (MAE) replaces the ``probability of failure'' as a performance measure: in this setting, Equation~\ref{eq:one_trajectory} is the one showing an accuracy up to 100 times higher than VARS-TO.

Nevertheless, the advantage of VARS-TO over Sobol'-based indices is still remarkable if the goal is to rank parameters, and suggests that VARS-TO should be the sensitivity measure of choice if 1) the model under analysis is known to be fully additive before its execution, \emph{and} 2) computational efficiency is a priority. However, this condition is unlikely to apply to models of the Earth and Environmental domain, either because they encompass multiplicative terms and exponentials or because their mathematical complexity prevents the analyst from knowing their behavior before running the simulations. 

The assertion that VARS-TO is at least 20 times more efficient than Sobol'-based total-order indices is not confirmed by our results. VARS-TO only very slightly outperforms one of the most accurate Sobol' total-order estimators, that of \textcite{Jansen1999}, when the number of model runs per model input is very small. However, it comes second to Jansen at increasing dimensionalities and in overall performance. Such results have been obtained after randomiaing the benchmark settings, thus creating a set of sensitivity problems much wider than those represented by the HYMOD and MESH models, and by simultaneously examining the internal uncertainties of VARS-TO ($N_{star},h$). Its sampling design makes it especially vulnerable to the high-order interactions between the sampling method ($\tau$), the number of stars ($N_{star}$), the function dimensionality ($k$), the distance between pairs ($h$) and the underlying distribution of the model inputs ($\phi$), especially if they follow a normal distribution.

VARS nonetheless represents a relevant addition to the family of sensitivity analysis methods, with the additional merit of having been developed to appraise the response surface of a model. Furthermore, the conceptual framework of VARS comes with a software described as \say{next-generation} by \textcite{Razavi2019}. Time will tell whether VARS ends up unseating Sobol'-based indices as the recommended best practice in sensitivity analysis. 

\section{Acknowledgements} 
This work has been funded by the European Commission (Marie Sk\l{}odowska-Curie Global Fellowship, grant number 792178 to A.P.). 

\section{Code availability}
Fully documented code is freely available in \textcite{Puyo} and in GitHub (\url{https://github.com/arnaldpuy/VARS_paper}).

\section{Data availability}
A \texttt{.csv} file of the studies citing VARS as of November 2020 can be found in GitHub (\url{https://github.com/arnaldpuy/VARS_paper}.

\printbibliography

\end{document}


\pagenumbering{arabic}
\date{}
\title{Is VARS more intuitive and efficient than Sobol' indices? \\ \vspace{3mm} \large{Supplementary Materials}}

\author[1,2]{Arnald Puy\thanks{Corresponding author}}
\author[3]{Samuele Lo Piano}
\author[2,3]{Andrea Saltelli}

\affil[1]{\footnotesize{\textit{Department of Ecology and Evolutionary Biology, M31 Guyot Hall, Princeton University, New Jersey 08544, USA. E-Mail: \href{mailto:apuy@princeton.edu}{apuy@princeton.edu}}}}

\affil[2]{\footnotesize{\textit{Centre for the Study of the Sciences and the Humanities (SVT), University of Bergen, Parkveien 9, PB 7805, 5020 Bergen, Norway.}}}

\affil[3]{\footnotesize{\textit{Open Evidence, Universitat Oberta de Catalunya, Edifici 22@, Universitat Oberta de Catalunya, 08018 Barcelona, Spain.}}}

\maketitle

\tableofcontents

\newpage

\beginsupplement

\section{The issue of intuitivity and importance}

\subsection{Formulae in Figure 1}
\label{sec:formulae}

Figure 1a:

\begin{equation}
\begin{aligned}
 f_1(x) & = x^2 \\
 f_2(x) & = \begin{cases} -x, & x < 0 \\ x, & \geq 0 \end{cases} \\
 f_3(x) & = \begin{cases} -(x+1)^2, & x > 0 \\ -(x-1)^2, & x\geq 0 \end{cases}
\end{aligned}
\end{equation}

Figure 1b:

\begin{equation}
\begin{aligned}
f_1(x) & =1.11x^2 \\
f_2(x) & = 2 - 0.2 \cos(7\pi x)
\end{aligned}
\end{equation}

Figure 1c:

\begin{equation}
\begin{aligned}
f_1(x) & =x \\
f_2(x) & = (-1)^ {\lvert 4x \rvert}\left [0.125-mod(x, 0.25) \right ] + 0.125 \\
f_3(x) & = (-1)^ {\lvert 32x \rvert}\left [0.0325-mod(x, 0.0325) \right ] + 0.0325
\end{aligned}
\end{equation}

Figure 1d:

\begin{equation}
\begin{aligned}
g_1(x_1) & =-\sin (\pi x_1) - 0.3 \sin (3.33\pi x_1) \\
g_2(x_2) & = -0.76 \sin \left [ \pi (x_2 - 0.2) \right ] - 0.315 \\
g_3(x_3) & = -0.12 \sin \left [1.05\pi(x_3 - 0.2) \right ] - 0.02 \sin (95.24 \pi x_3)-0.96 \\
g_4(x_4) & = -0.12 \sin \left [ 1.05 \pi(x_4 - 0.2)\right] - 0.96 \\
g_5(x_5) & = -0.05 \sin \left[ \pi ( x_5 - 0.2) \right ] - 1.02 \\
g_6(x_6) & = -1.08 \\
y & =f \left [ g_1(x_1) + g_2(x_2) +...+g_6(x_6) \right ]
\end{aligned}
\label{eq:six_dimensional}
\end{equation}

\section{The issue of efficiency}

\subsection{The case of the six-dimensional response surface model}

The following proves that if a function is of the additive type, its first-order index $S_i$ can be computed very easily with a single trajectory along the $x$ axis, irrespective of its position in the space of $\bm{x}_{\sim i}$.

\vspace{5mm}

\textit{Proof}:
If the function is of the type 

\begin{equation}\label {F_def}
F=f(x_i)+g(\bm{x}_{\sim i})\,,
\end{equation}
and we wish to compute 

\begin{equation}\label {Si}
S_i=\frac{V_{x_i}(E_{\bm{x}_{\sim i}}(F|x_i))}{V(F)}\,,
\end{equation}

$V_{x_i}(E_{\bm{x}_{\sim i}}(F|x_i))$ can be written as 

\begin{equation}\label {VofE}
V_{x_i}(E_{\bm{x}_{\sim i}}(F|x_i))=E_{x_i}(E_{\bm{x}_{\sim i}}(F|x_i))^2-E_{x_i}^2(E_{\bm{x}_{\sim i}}(F|x_i))\,.
\end{equation}

The inner sum $E_{\bm{x}_{\sim i}}(F|x_i)$ can be written in terms of $f$ and $g$:

\begin{equation}\label {inner}
E_{\bm{x}_{\sim i}}(F|x_i)=E_{\bm{x}_{\sim i}}(f(x_i)+g(\bm{x}_{\sim i})|x_i)=f(x_i)+\hat{g}\,,
\end{equation}

where we indicate as $\hat{g}$ the average of $g(\bm{x}_{\sim i})$. We now compute the two terms in Equation~\ref{VofE}; the first term is:

\begin{equation}\label {VofE1}
E_{x_i}(E_{\bm{x}_{\sim i}}(F|x_i))^2=E_{x_i}(f(x_i)+\hat{g}))^2=E_{x_i}(f(x_i))^2+\hat{g}^2+2\hat{g}E_{x_i}(f(x_i)) \,,
\end{equation}

and the second term is:

\begin{equation}\label {VofE2}
E_{x_i}^2(E_{\bm{x}_{\sim i}}(F|x_i))=E_{x_i}^2(f(x_i)+\hat{g})=E_{x_i}^2(f(x_i))+\hat{g}^2+2\hat{g}E_{x_i}(f(x_i))\,.
\end{equation}

Substituting Equations~\ref{VofE1}--\ref{VofE2} into \ref{VofE} leads to

\begin{equation}
V_{x_i}(E_{\bm{x}_{\sim i}}(F|x_i))=E_{x_i}(f(x_i))^2-E_{x_i}^2(f(x_i)\,.
\label{eq:VofEfin}
\end{equation}
 
which proves that if function $F$ is additive in factor $x_i$ its sensitivity index only depends upon $f$.  

\subsubsection{Analytical values}

The six-dimensional model (Equation~\ref{eq:six_dimensional}) has the following analytic values (Table~\ref{tab:analytical}):

\begin{table} [ht]
\begin{center}
{\scriptsize
\begin{tabular} {l l l l}
\toprule
  & Function & $V(f_i(x_i))$ \\
\hline
  & $g_1(x_1)$  & 0.0972 \\
  & $g_2(x_2)$  &  0.136\\
  & $g_3(x_3)$  &  0.00358\\
  & $g_4(x_4)$  &  0.00301 \\
  & $g_5(x_5)$  &  0.000587 \\
  & $g_6(x_6)$  &  0 \\
\bottomrule  
\end{tabular}
}
\end{center}
\caption {$V(f_i(x_i))$ for Equation~\ref{eq:six_dimensional}.} \label{tab:analytical}
\end{table}

All functions are covered by the general formula corresponding to $g_3(x_3)$: 

\begin{equation} 
a\sin(b(x+e))+c\sin(dx)+f
\label{eq:g3}
\end{equation}

To make an example, $g_1(x_1)$ corresponds to \ref{eq:g3} when $f$ is set to zero, while for $g_2(x_2)$ $c$ is zero. Thus we only offer the analytic expression for $g_3(x_3)$ which is obtained as from Equation~\ref{eq:VofEfin}:

\begin{equation}\label{Eof2} 
\left.\begin{aligned}
E_{x_3}\left(g_3(x_3) \right)^2 =   \frac{1}{\pi}    \Bigg| & \frac{a^2}{2b}\bigg(b(x+e)-\frac{1}{2}\sin\big(2b(x+e)\big)\bigg)  +   \\
    &   \frac{c^2}{2d}\bigg(dx-\frac{1}{2}\sin\big(2dx)\big)\bigg)  +   \\
    &   f^2x-   \\
    &   \frac{2af}{b} \cos(b(x+e))-   \\
    &   \frac{2cf}{d} \cos(dx)+   \\
    &   2ac \cos(de) \bigg ( \frac{\sin[(b-d)x]}{2(b-d)}-\frac{\sin[(b+d)x]}{2(b+d)} \bigg)  +   \\
    &   2ac \sin(de)\bigg ( -\frac{\cos[(b-d)x]}{2(b-d)}-\frac{\cos[(b+d)x]}{2(b+d)} \bigg)  \Bigg|_0^{\pi}  \,,
\end{aligned}\right.
\end{equation}
and

\begin{equation}\label{E2of} 
\left.\begin{aligned}
E_{x_3}^2\left(g_3(x_3) \right) =   \frac{1}{pi}    \Bigg| & \frac{a}{b} \cos \big(b(x+e) \big) +   \\
    & \frac{c}{d}\cos (dx) +    fx  \Bigg|_0^{pi} \,,
\end{aligned}\right.
\end{equation}
to give 

\begin{equation}\label{fin} 
\begin{aligned}
S_{x_3}=\frac { E_{x_3}\left [ g_3(x_3) \right ] ^2  -E_{x_3}^2\left [ g_3(x_3) \right ] }{V(y)}\,.
\end{aligned}
\end{equation}

\subsection{The case of the HYMOD and MESH models}

\subsubsection{The rationale of our experiment}

Here we describe the conceptual and methodological approach behind our assessment of the performance of VARS-TO \cite{Razavi2016a, Razavi2016b} and the \textcite{Jansen1999} estimator. We basically followed \textcite{Becker2019, Puyj}, to which we direct the reader for more detailed information. The main rationale is the following:

\begin{enumerate}
\item The performance of sensitivity indices is usually benchmarked with a handful of test functions and across different sample sizes \cite{Janon2014, Azzini2019, Saltelli2010a, LoPiano2020, Razavi2016a, Razavi2016b}. However, there are other factors that might also condition the behavior of sensitivity indices besides the model and the number of allocated runs: 

\begin{enumerate}
\item The sampling method used to create the sample matrix (i.e. random numbers, Quasi-random numbers) \cite{Kucherenko2011}.
\item The distribution of the model inputs \cite{Shin2013, Paleari2016}.
\item The dimensionality of the model.
\item Its degree of non-additivity.
\item The performance measure used (i.e. whether we assess how well the estimated indices approach the "true" indices, how well the estimator ranks the parameters, or how well the estimator ranks the most important parameters only) \cite{Becker2019, Puyj}.
\item The design parameters  (i.e. the number of stars $N_{star}$ or the value of $h$ for VARS-TO).
\end{enumerate}

\item By treating these factors as uncertain parameters we can check the accuracy of sensitivity indices in a comparatively much larger range of sensitivity settings. This provides a much thorough picture of their behavior and a clearer assessment of their advantages and limitations \cite{Puyj}. In a sense, this approach simulates what happens when all the forking paths leading to the computation of a sensitivity index are walked at once \cite{Puy2020, Borges1941}.

\item To follow this approach, we described these uncertain parameters with probability distributions based on the literature available (Fig.~\ref{fig:tree}):

\begin{enumerate} 
\item $N_{star}$ was defined based on the range of star centers used by \textcite{Razavi2016a, Razavi2016b} in their examples.
\item $h$ was described with the values used by \textcite{Becker2019,Razavi2016a, Razavi2016b}.
\item $k$ was defined to explore the behavior of VARS-TO and Jansen at both small and medium dimensionalities. For $k>50$, see \textcite{Puyj}.
\item $\epsilon$ sets the seed of the random numbers required to define the test function. Our test function follows the metafunction approach of \textcite{Becker2019}: it allows test functions to be generated by randomly combining $p$ univariate functions in a multivariate function of dimension $k$. Our metafunction includes cubic, discontinuous, exponential, inverse, linear, non-monotonic, periodic, quadratic, trigonometric functions, as well as a no-effect function (Fig.~\ref{fig:metafunction}). See \textcite{Becker2019, Puyj} for more details.
\item $\tau$ was described to check the performance of VARS-TO and Jansen when the base sample matrix is constructed with random numbers ($\tau=1$) or with \textcite{Sobol1967, Sobol1976} Quasi-Random Numbers ($\tau=2$).
\item $\phi$ aimed at observing how the estimators behave under uniform, normal, skewed and random distributions (Fig.~\ref{fig:distributions}).
\item $k_2$ was described to randomly activate between 50\% and 100\% of the existing pairwise interactions of the test function.
\item $k_3$ was described to randomly activate between 30\% and 100\% of the three-wise interactions of the test function
\item $\delta$ checked how the estimators behave when the performance measure varies. Let $\bm{T}$ and $\bm{\hat{T}}$ be vectors with the \say{true} and the estimated sensitivity indices respectively.  We checked how well $\bm{\hat{T}}$ correlated with $\bm{T}$ ($\delta=1$) , how well the ranks of $\bm{\hat{T}}$ correlated with those of $\bm{T}$ ($\delta=2$), and how well the most important ranks of $\bm{\hat{T}}$ correlated with those of $\bm{T}$ ($\delta=3$).
\end{enumerate}
\end{enumerate}

\subsubsection{The simulations}

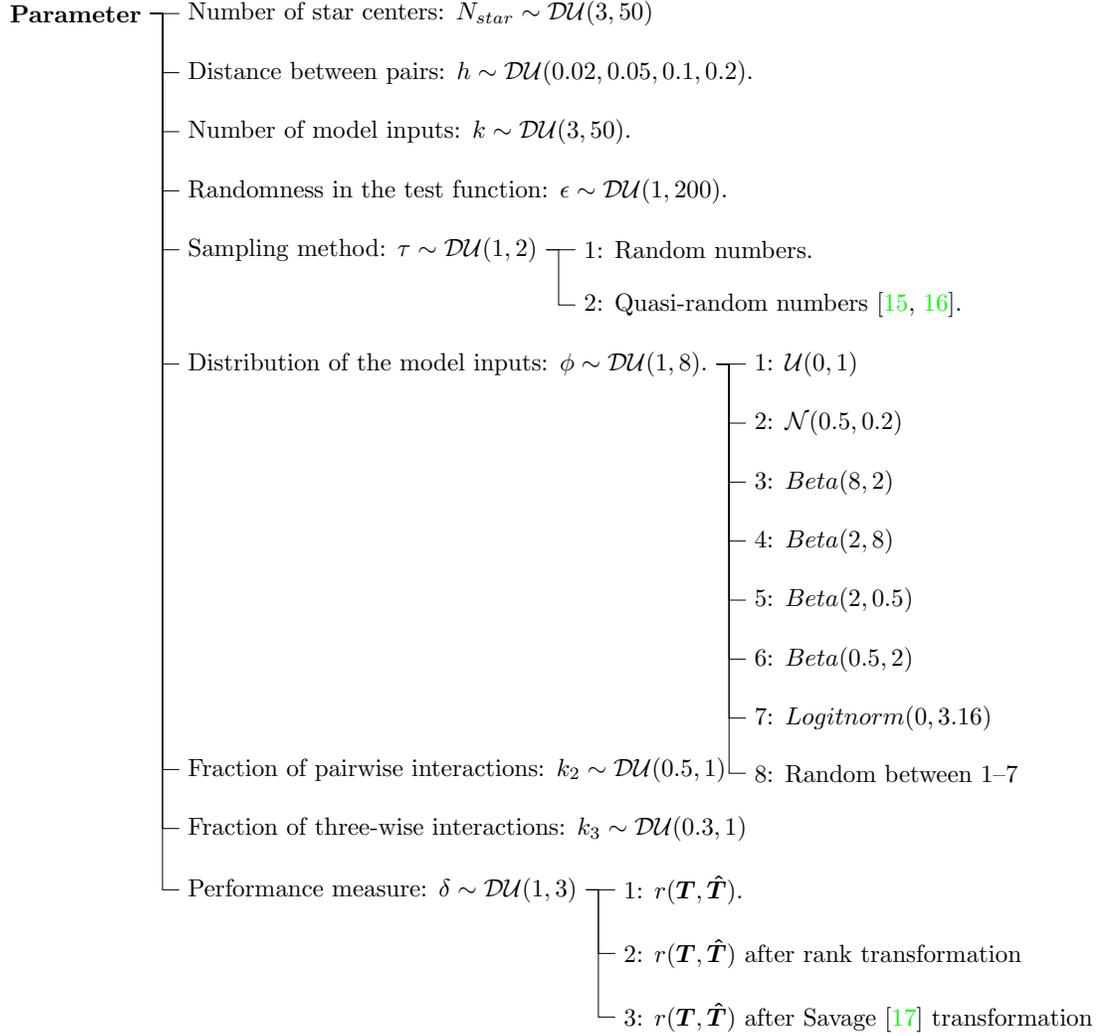
\begin{figure}[ht]
\begin{forest}
  for tree={
    grow'=0,
    parent anchor=children,
    child anchor=parent,
    anchor=parent,
    if n children=0{folder}{},
    edge path'={(!u.parent anchor) -- ++(5pt,0) |- (.child anchor)},
  },
  where n=1{
    calign with current edge
  }{},
   [\textbf{Parameter} 
        [{Number of star centers: $N_{star}\sim\mathcal{DU}(3, 50)$}]
        [{Distance between pairs: $h\sim\mathcal{DU}(0.02, 0.05, 0.1, 0.2)$}.]
        [{Number of model inputs: $k\sim\mathcal{DU}(3,50)$}.]
        [{Randomness in the test function: $\epsilon\sim\mathcal{DU}(1,200)$}.]
        [{Sampling method: $\tau\sim\mathcal{DU}(1, 2)$}
         [{1: Random numbers. } ]
         [{2: Quasi-random numbers \cite{Sobol1967, Sobol1976}. } ]
         ]
       [{Distribution of the model inputs: $\phi\sim\mathcal{DU}(1,8)$}.
        [{1: $\mathcal{U}(0, 1)$ } ]
        [{2: $\mathcal{N}(0.5, 0.2)$ } ]
        [{3: $Beta(8, 2)$ } ]
        [{4: $Beta(2, 8)$ } ]
        [{5: $Beta(2, 0.5)$ } ]
        [{6: $Beta(0.5, 2)$ } ]
        [{7: $Logitnorm(0, 3.16)$ } ]
        [{8: Random between 1--7 } ]
       ]
       [{Fraction of pairwise interactions: $k_2\sim\mathcal{DU}(0.5, 1)$}]
        [{Fraction of three-wise interactions: $k_3\sim\mathcal{DU}(0.3, 1)$}]
        [{Performance measure: $\delta\sim\mathcal{DU}(1, 3)$}
          [{1: $r(\bm{T}, \bm{\hat{T}})$.} ]
         [{2: $r(\bm{T}, \bm{\hat{T}})$ after rank transformation} ]
        [{3: $r(\bm{T}, \bm{\hat{T}})$ after \textcite{Savage1956} transformation} ]
        ]
    ]
\end{forest}
\caption{Tree diagram with the uncertain parameters and their levels.}
\label{fig:tree}
\end{figure}

\begin{figure}[ht]
\centering
\includegraphics[keepaspectratio]{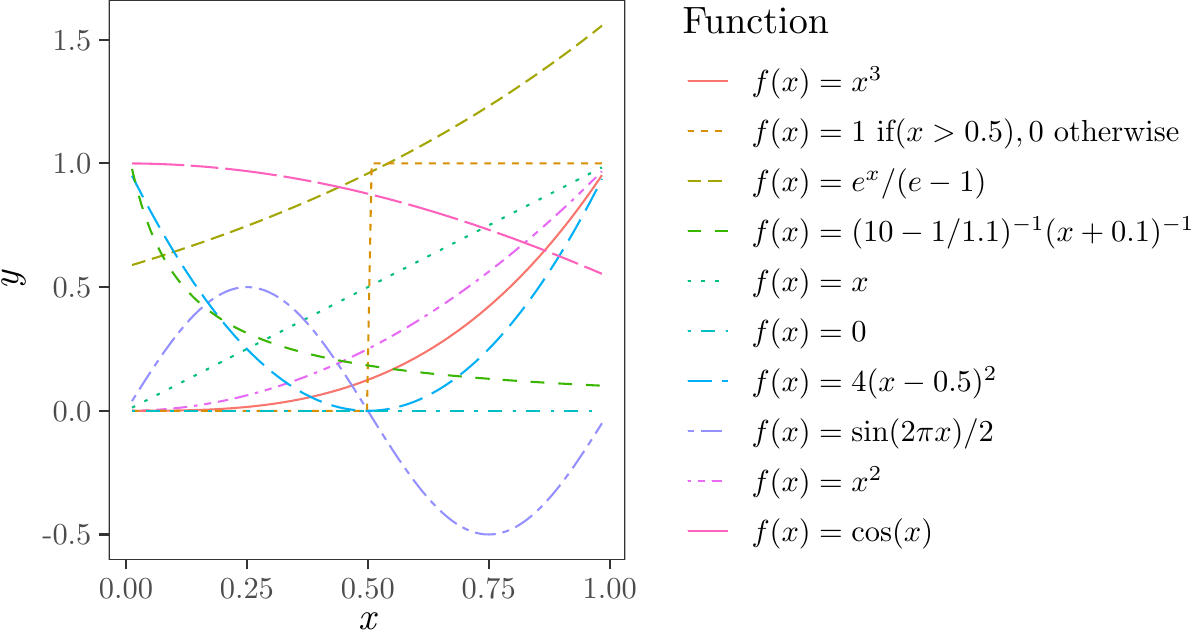}
\caption{Univariate functions included in the metafunction.}
\label{fig:metafunction}
\end{figure}

\begin{figure}[ht]
\centering
\includegraphics[keepaspectratio]{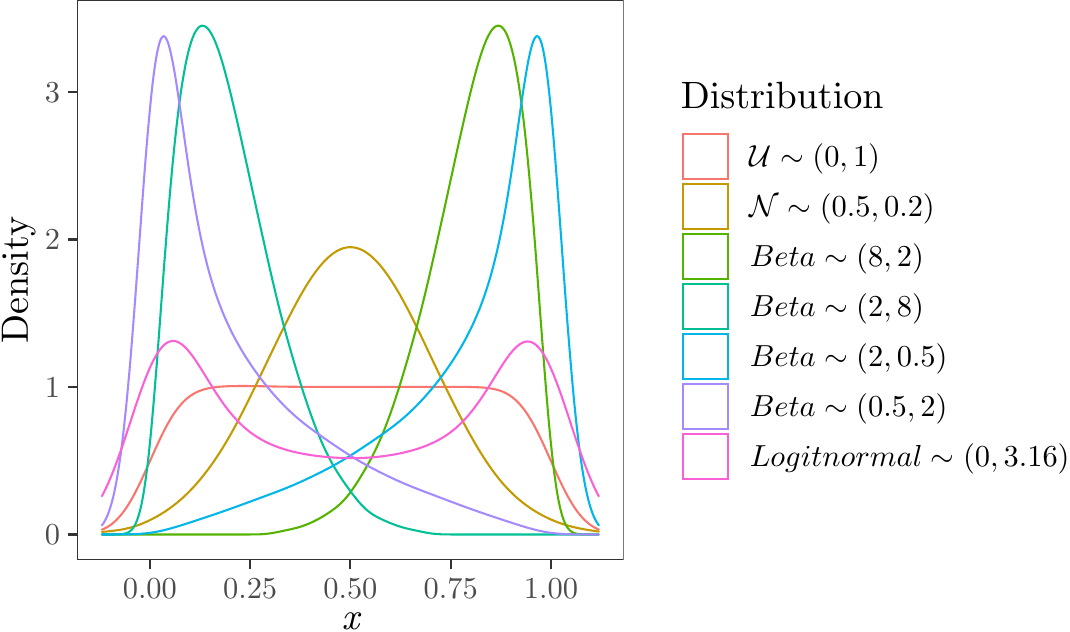}
\caption{Probability distributions included in $\phi$.}
\label{fig:distributions}
\end{figure}

\begin{figure}[ht]
\centering
\includegraphics[keepaspectratio]{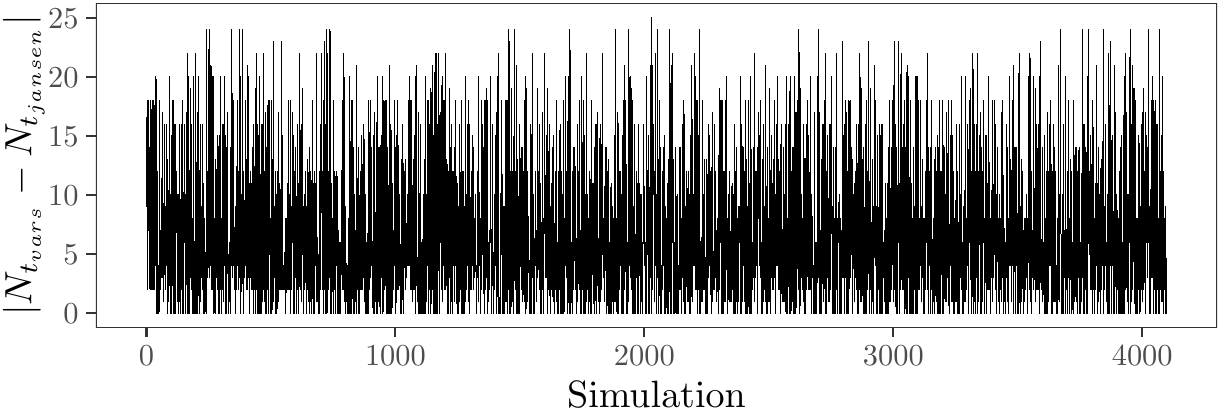}
\caption{Line plot displaying the absolute difference between the total number of model runs allocated to Jansen and the total number of runs allocated to VARS-TO in each simulation. }
\label{fig:sample_size}
\end{figure}

\begin{enumerate}
\item We created a $(2^{12}, 2k)$ sample matrix using Sobol' Quasi-Random Numbers, where the $k$ leftmost columns are allocated to an $\bm{A}$ matrix and the $k$ leftmost columns to a $\bm{B}$ matrix. In these matrices, each row is a sample point and each column a factor input, distributed as in Fig.~\ref{fig:tree}. We then added $k$ $(2^{12}, k)$ $\bm{A}_{B}^{(i)}$ matrices, where all columns come from $\bm{A}$ except the $i$-th, which comes from $\bm{B}$. This was needed to compute the Sobol' indices of the parameters in Fig.~\ref{fig:tree}.

\item Our model ran rowwise over the $\bm{A}$, $\bm{B}$ and $\bm{A}_{B}^{(i)}$ matrices for a total computational cost of $N_t=2^{12}(k+2)=45056$, $k=9$. For $v=1,2,...,N_t$ rows, our model does the following:

\begin{enumerate}
\item It designs three sample matrices using either random numbers ($\tau_v=1$) or quasi-random numbers ($\tau_v=2$):
\begin{enumerate}
\item A sample matrix for VARS-TO, $N_{t_{vars_{v}}}=N_{star_{v}} \left [ k_v ( \frac{1}{\Delta h_v} - 1) + 1 \right ]$.
\item A sample matrix for Jansen, $N_{t_{jansen_{v}}}=N_v(k_v+1)$, where $N_v= \frac{N_{t_{vars_{v}}}}{k_v+1}$ and $N_v=2$ if $2 > \frac{N_{t_{vars_{v}}}}{k_v+1}$. This forced  all comparisons between VARS-TO and Jansen to be conducted on an identical or almost identical total number of runs. Fig.~\ref{fig:sample_size} shows that the total number of model runs allocated to each estimator in each simulation differed mostly by 1-10 model runs, and that the largest difference was of 25 model runs only.
\item A large sample matrix to compute $\bm{T}_v$, $N_t=2^{12}(k_v + 1)$.
\end{enumerate}

\item It transforms the model inputs of all three matrices into the probability distribution set by $\phi_v$.

\item It applies the metafunction to all three matrices simultaneously, with its functional form, degree of active and third-order effects defined by $\epsilon_v$, $k_{2_{v}}$ and $k_{3_{v}}$ respectively.

\item It computes the estimated sensitivity indices $\bm{\hat{T}}_v$ for VARS-TO and Jansen, and the \say{true} sensitivity indices $\bm{T}_v$ from the large sample matrix (see above).

\item It assesses the performance of VARS-TO and Jansen according to $\delta_v$: if $\delta=1$, it checked how well $\bm{\hat{T}}_v$ correlated with $\bm{T}_v$; if $\delta=2$, how well the ranks of $\bm{\hat{T}}_v$ correlate with those of $\bm{T}_v$; and if $\delta=3$, how well the ranks of the most important parameters of $\bm{\hat{T}}_v$ correlated with those of $\bm{T}_v$. The resulting model output was the correlation coefficient $r_v$.
\end{enumerate}
\end{enumerate}

\begin{figure}[ht]
\centering
\includegraphics[keepaspectratio]{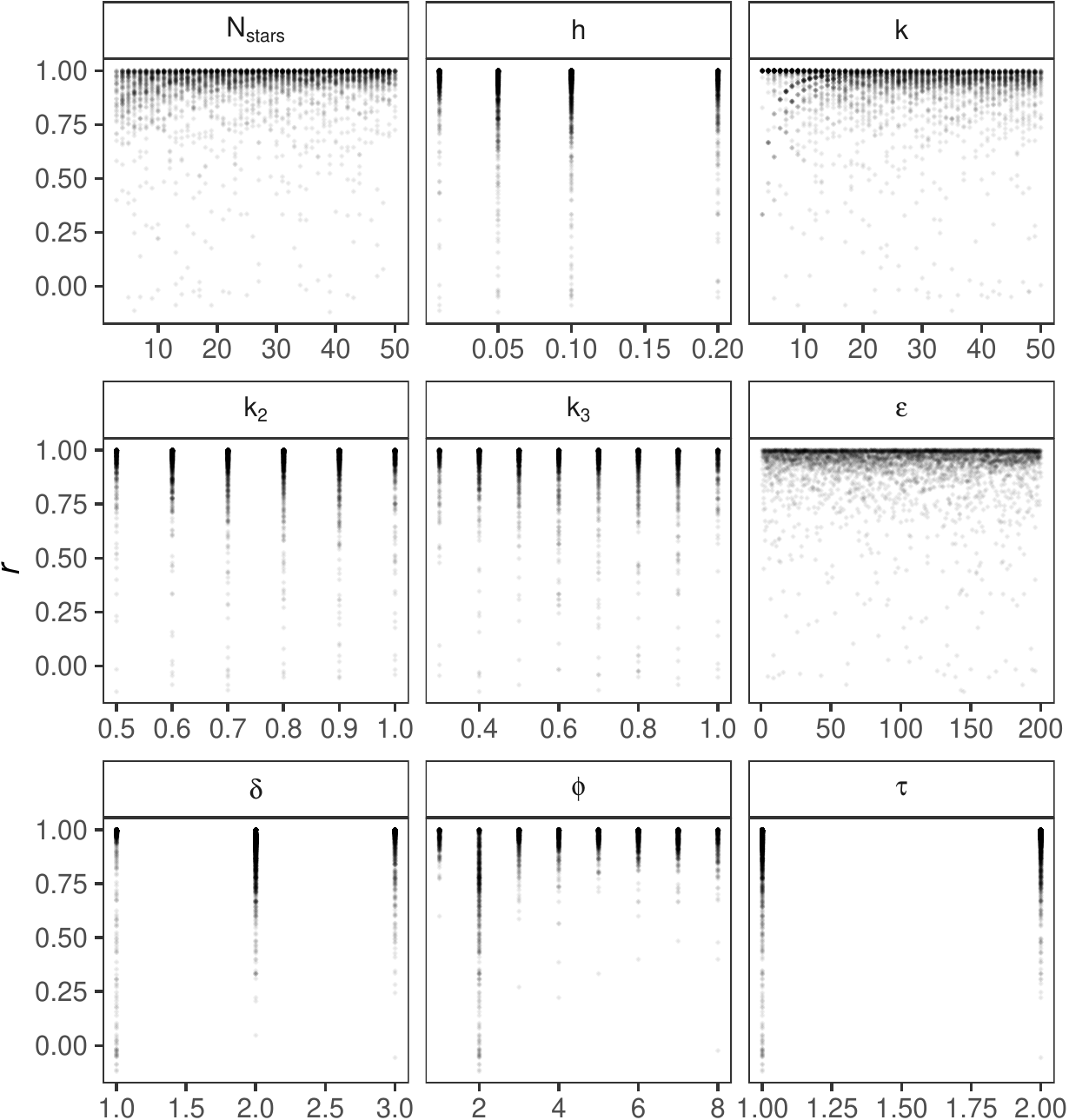}
\caption{Scatterplots of the model inputs against the model output.}
\label{fig:scatter}
\end{figure}

\clearpage

\printbibliography